\documentclass[manuscript]{aastex}

\usepackage{euscript,amssymb,amsmath}

\newcommand{\be}[1]{\begin{equation}\label{#1}}
\newcommand{\ee}{\end{equation}}
\newcommand{\ba}[1]{\begin{eqnarray}\label{#1}}
\newcommand{\ea}{\end{eqnarray}}
\newcommand{\rf}[1]{(\ref{#1})}
\newcommand{\nn}{\nonumber}

\slugcomment{Submitted to The Astrophysical Journal}

\shorttitle{On the relation of standard and helical magnetorotational instability}
\shortauthors{KS}

\begin{document}


\title{On the relation of standard and helical magnetorotational instability}


\author{Oleg N. Kirillov} 
\affil{Dynamics and Vibrations Group, Technical University of Darmstadt,
Hochschulstr. 1, 64289 Darmstadt, Germany}
\email{kirillov@dyn.tu-darmstadt.de}

\author{Frank Stefani} 
\affil{Forschungszentrum Dresden-Rossendorf, P.O. Box 510119, D-01314 Dresden, Germany}
\email{f.stefani@fzd.de}




\begin{abstract}
The magnetorotational instability (MRI) plays a crucial
role for cosmic structure formation by enabling turbulence
in Keplerian disks which would be otherwise hydrodynamically
stable. With particular focus on MRI experiments with liquid metals,
which have small magnetic Prandtl numbers,
it has been shown that
the helical version
of this instability (HMRI) has a scaling behaviour that is
quite different from that of the standard MRI (SMRI). We discuss the
relation of HMRI to SMRI  by exploring various parameter
dependencies. We identify the mechanism of transfer of instability     
between modes through a spectral exceptional point that explains
both the transition from a stationary instability (SMRI)              
to an unstable travelling wave (HMRI) and the excitation                 
of HMRI in the inductionless limit. For certain parameter regions
we find new islands of the HMRI.
\end{abstract}





\section{Introduction}

The magnetorotational instability (MRI) \citep{B09} is considered as
the main candidate to solve the long-standing puzzle of
how stars and black holes are fed by the
accretion disks surrounding them. The central problem is that
these accretion disks typically rotate according to Kepler's
law, $\Omega(r) \sim r^{-3/2}$, which results in an
angular momentum $r^2 \Omega(r) \sim r^{1/2}$.
Hence, they fulfill Rayleigh's criterion stating that
rotating flows with radially increasing angular momentum
are hydrodynamically stable, at least in the linear
sense. Such stable, non-turbulent disks would not allow
the outward directed angular momentum transport that is
necessary for the infalling disk matter to accrete into
the central object.

In their seminal paper of 1991 \cite{BH91}
Balbus and Hawley had highlighted the
key role of the magnetorotational instability (MRI)
in explaining turbulence and angular momentum
transport in accretion disks around stars and black holes.
They had shown that a weak, externally
applied magnetic field serves
only as a trigger for the instability that
actually taps into the rotational energy of the flow.
This is quite in contrast to current-induced instabilities,
e.g. the Tayler instability \cite{T73}, which draw
their energy (at least
partly) from
the electric currents in the fluid.

Soon after the paper by Balbus and Hawley it became clear that the
principle mechanism of the MRI had already been revealed three
decades earlier by Velikhov \cite{V59} and Chandrasekhar \cite{C60}.
Actually, they had investigated the destabilizing action of an
external magnetic field for the classical Taylor-Couette (TC) flow
between two concentric, rotating cylinders rather than for Keplerian
rotation profile. This is, however, not a crucial difference since a
TC flow can be made very close to  a Keplerian one simply by
adjusting the ratio of rotation rates of the inner and the outer
cylinder.

The MRI in flows between rotating walls has attracted             
renewed interest during the last decade, mainly motivated
by the increasing efforts to investigate MRI in the
laboratory \cite{AIP04,SGG08}.                                    
A first interesting experimental result
was obtained in a spherical Couette flow
of liquid sodium \cite{SMTHDHAL04}. The authors                    
observed correlated modes of velocity and magnetic field
perturbation in a parameter region
which is quite typical for MRI. It                               
must be noted, however, that the                                  
background state in this spherical Couette experiments
was already fully turbulent, so that the
original goal that the MRI would destabilize an
otherwise stable flow was not met.
At Princeton University work is
going on to identify MRI in a TC experiment
with liquid gallium,
and first encouraging results, including the
observation of
non-axisymmetric Magneto-Coriolis waves,
have been obtained \cite{N08, N09}.

Both experiments had been designed to
investigate the standard
version of MRI (SMRI) with only a vertical
magnetic field being applied. In this case, the
azimuthal magnetic field (which is an essential ingredient of the
MRI mode) must be  produced from the vertical field
by induction effects, which are proportional to the magnetic
Reynolds number (${\rm Rm}$) of the flow. ${\rm Rm}$, in turn, is
proportional to the hydrodynamic Reynolds number according to
${\rm Rm}={\rm Pm} {\rm Re}$, where the magnetic Prandtl
number ${\rm Pm}=\nu/\eta$
is the ratio of viscosity $\nu$ to magnetic
diffusivity $\eta=1/\mu_0 \sigma$.
For liquid metals ${\rm Pm}$ is typically in the
range $10^{-6}...10^{-5}$.
Therefore, in order to achieve ${\rm Rm} \sim 1$, we need
${\rm Re}\sim 10^5...10^6$, and wall-constrained flows
(in contrast to wall-free Keplerian flows)
with such high ${\rm Re}$ are usually turbulent,
whatever the linear stability analysis might tell
(see, however, \cite{JINATURE}).                                
This is the point which makes SMRI experiments, and their
interpretation, so cumbersome.

One might ask, however, why not to
substitute the induction of the
necessary azimuthal magnetic field
component of the MRI mode
by simply externally applying this component               
as a part of the base configuration.
Indeed, it was shown \cite{HR05,RUEDIGER2005}                   
that the resulting ''helical MRI'' (HMRI),
as we now call it, is then possible at                      
far smaller Reynolds numbers and magnetic field amplitudes
than SMRI, making HMRI an ideal playground for liquid
metal experiments.

First experimental evidence for HMRI
was obtained in 2006 at the liquid metal facility
PROMISE ({\it P\/}otsdam {\it RO\/}ssendorf
{\it M\/}agnetic {\it I\/}n{\it S\/}tability {\it E\/}xperiment)
which is basically  a Taylor-Couette (TC) cell made of
concentric rotating
copper walls, filled with  GaInSn (a eutectic
which is liquid at room temperatures).
In \cite{SGGRSSH06,RHSGGR06,SGGRSH07,SGGSRH08} it
was shown that the
HMRI travelling wave appears only in
the predicted finite window
of the magnetic field intensity, with a
frequency of the
travelling wave that was also in good accordance with
numerical simulations. Results of a significantly
improved experiment (PROMISE 2) with
strongly reduced Ekman pumping at the end-caps
were published recently \cite{SGGSRH09,SGGHPRS09}.

The connection of SMRI and HMRI is presently
under intense
debate \cite{LGHJ06,RH07,PGG07,LV07,LGJ07,S07,RS08,L09,PG09}.
The first essential
point to note here is that HMRI and SMRI are               
connected. Indeed, Fig. 1 in
\cite{HR05} shows that there is a
continuous and monotonic transition from HMRI
to SMRI
when ${\rm Re}$ and the magnetic field
strength are
increased simultaneously.

A second remarkable property of HMRI for small
${\rm Pm}$ (which has been
coined ``inductionless MRI''), was
clearly worked out in
\cite{PGG07}. It is the apparent paradox that a
magnetic field is able to trigger an instability
although the total energy dissipation of the system is
larger than without this field.

The relevance of HMRI for Keplerian flows has               
been seriously put into question in \cite{LGHJ06}.
Using a local WKB analysis in the small-gap approximation,
the authors had shown that the HMRI works only for comparably
steep rotation profiles (i.e. slightly above
the Rayleigh line)
and disappears for profiles
as flat as the Keplerian one.
This result has been confirmed by Lakhin and
Velikhov \cite{LV07}
and R\"udiger and Schultz \cite{RS08}.

However, this disappointing result was soon relativized in
\cite{RH07} by solving the global eigenvalue equation for HMRI with
electrical boundary conditions. It turned out that HMRI re-appears
again for Keplerian flows provided that at least one radial boundary is    
highly conducting. A similar discrepancy between local and global
results is well known for the so-called stratorotational instability
(SRI) \cite{DMNRHZ05} for which the existence of reflecting
boundaries appears necessary for the instability to work \cite{U06}.
This artificial demand is of course a much stronger argument against
the working of SRI than the necessity  of one conducting boundary is
for the working of HMRI: considering, i.e., the colder outer parts
of accretion disks, then the inner part can indeed be considered as
a good conductor \cite{BALBUS2008}.                                        

Another argument that has been put forward against the
relevance of HMRI for thin
accretion disks is the
necessity for a large ratio of toroidal to poloidal
magnetic fields \cite{L08}.

A further complication for applying HMRI to the real world     
is the fact that it appears in form of a travelling wave.
The crucial point here is that
monochromatic waves are typically not able                       
to fulfill the axial boundary conditions at the
ends of the considered region.
To fulfill them,  one has to consider wave packets.
Only wave packets with vanishing group velocity
will remain in the finite length system.
Typically, the onset of this {\it absolute instability},
characterized by
a zero growth rate {\it and} a zero group velocity, is harder
to achieve  than the
{\it convective instability} of a monochromatic wave
with zero growth rate.
A comprehensive analysis of the relation of convective and        
absolute instability for HMRI
can be found in \cite{PG09}.
From the extrapolation of the results
of this paper it seems that
Keplerian rotation profiles (with conducting boundaries)          
are indeed
absolutely HMRI-unstable, but a final solution to
this puzzle is still elusive.

In the present paper, we step back from those                 
important consideration of absolute and global instabilities
and focus again on the local WKB method
by considering the
dispersion relation of MRI which had been                           
derived and analyzed in \cite{LGHJ06,LV07,RS08}.
In spite of these former investigations,                         
we feel that some points still need further clarification.      
This concerns a careful application of the                     
Bilharz stability criterion
as well as some further parameter
dependencies, in particular the dependence for small
but finite magnetic Prandtl numbers.
It also concerns the question in which sense the HMRI
can be considered as a {\it dissipation induced instability}
which
is quite common in many areas of physics
\cite{KM07,Ki07}.

To make the paper self-contained, we will start with a
re-derivation of the
dispersion relation in two forms which explicitly contain
the relevant frequencies or the
dimensionless parameters, respectively.

Then we will study the peculiar
relation of SMRI and HMRI. As a main result of this paper
we will describe in detail the mechanism of transition
from SMRI to HMRI through a spectral exceptional point
which appears at finite but small ${\rm Pm}$.
This provides  a natural explanation for the         
continuous and monotonic connection between SMRI          
(a destabilized slow magneto-Coriolis wave)                          
and HMRI (a weakly destabilized inertial oscillation).     
In addition to this, for high Reynolds numbers we will            
identify a second scenario for HMRI which leads to             
new islands of instability at small but finite values          
of $\rm Pm$.

\section{Mathematical setting}

In order to make the paper self-contained,
we will re-derive in this section the dispersion
relation for HMRI, including viscosity and resistivity
effects. Note
that this dispersion relation
was derived in various forms
and approximations by a number of authors
\cite{LGJ07,LV07,GHSS08}.

The standard set of non-linear equations of
dissipative incompressible magnetohydrodynamics
\citep{JGK01,G02,No02,LV07,RS08} consists of the Navier-Stokes equation for
the fluid velocity ${\bf u}$
\be{m1}
\frac{\partial {\bf u}}{\partial t}+({\bf u} \cdot \nabla) {\bf u} = - \frac{1}{\rho} \nabla \left(p+\frac{{\bf B}^2}{2\mu_0}\right)+\frac{1}{\mu_0 \rho}({\bf B}\cdot \nabla){\bf B}+\nu \nabla^2 {\bf u},
\ee
and of the induction equation for the magnetic field ${\bf B}$
\be{m2}
\frac{\partial {\bf B}}{\partial t}=\nabla \times ({\bf u} \times {\bf B}) + \eta \bigtriangledown^2 {\bf B},
\ee
where $p$ is the pressure, $\rho=const$
the density, $\nu=const$ the kinematic
viscosity,
$\eta=(\mu_0 \sigma)^{-1}$ the magnetic diffusivity, $\sigma$ the conductivity of the fluid,
and $\mu_0$
the magnetic permeability of free space.
Additionally, the mass continuity equation for incompressible flows
and the solenoidal condition for the magnetic induction yield
\be{m3}
\nabla \cdot {\bf u} = 0,\quad  \nabla \cdot {\bf B}=0.
\ee

We consider the rotational fluid flow in the gap between the radii
$R_1$ and $R_2>R_1$, with an imposed magnetic field sustained by
currents external to the fluid. The latter is important in order to
distinguish the MRI from other instabilities (i.e. the Tayler
instability for which electric currents are applied to the fluid).       
Introducing the cylindrical coordinates $(R, \phi, z)$ we consider
the stability of a steady-state background liquid flow with           
the angular
velocity profile $\Omega(R)$ in helical background magnetic field (a
magnetized Taylor-Couette flow)
\be{m4} {\bf u}_0=R\,\Omega(R)\,{\bf
e}_{\phi},\quad p=p_0(R), \quad {\bf B}_0=B_{\phi}^0(R){\bf
e}_{\phi}+B_z^0 {\bf e}_z,
\ee
with the azimuthal component
\be{m4a}
B_{\phi}^0(R)=\frac{\mu_0 I}{2 \pi R},                                    
\ee which can be thought as
being produced by an axial current $I$. The angular velocity profile      
of the background TC flow is \be{m4b} \Omega(R)=a+\frac{b}{R^2}, \ee
where $a$ and $b$ are arbitrary constants as in Taylor-Couette
experiments \cite{W99}. The centrifugal acceleration of the
background flow \rf{m4b} is compensated by the pressure gradient
\cite{JGK01} \be{m4c} R\Omega^2=\frac{1}{\rho}\frac{\partial
p_0}{\partial R}. \ee

\subsection{Linearization with respect to axisymmetric perturbations}

Throughout the paper we will restrict our interest to axisymmetric
perturbations ${\bf u}'={\bf u}'(R,z)$, ${\bf B}'={\bf B}'(R,z)$,
and ${p}'={p}'(R,z)$ about the stationary solution \rf{m4}-\rf{m4c},
keeping in mind that for                                                
strongly dominant azimuthal magnetic fields also non-axisymmetric
perturbations are possible \cite{HTR09}.

With the notation
\be{al0}
D_1=\partial_R\partial_R^{\dagger}+\partial_z^2,\quad
D_2=\partial_R^{\dagger}\partial_R+\partial_z^2,
\ee
where the differential operators are defined in \rf{l0},
the general linearized equations \rf{l1} derived in the

Appendix are simplified in the assumption of axisymmetric perturbations to
\ba{al1}
(\partial_t-\nu D_1) u_R'-2\Omega u_{\phi}'
&=&
-\frac{1}{\rho}\left[\partial_R p'+\frac{1}{\mu_0}\left(B_z^0{\partial_R}B_z' + B_{\phi}^0{\partial_R}B_{\phi}'\right)\right]
+
\frac{1}{\mu_0 \rho} \left( B_z^0 {\partial_z}B_R' -\frac{B_{\phi}^0}{R} B_{\phi}'\right),\nn\\
({\partial_t}-\nu D_1)u_{\phi}'+\frac{\kappa^2}{2\Omega}u_R'&=&
\frac{B_z^0}{\mu_0\rho} {\partial_z B_{\phi}'},\nn\\
({\partial_t}-\nu D_2)u_z'&=&
-\frac{1}{\rho}\left[{\partial_z p'}+\frac{1}{\mu_0}\left(B_z^0{\partial_z B_z'} + B_{\phi}^0{\partial_z B_{\phi}'}\right)\right]+\frac{B_z^0}{\mu_0\rho} {\partial_z B_z'},\nn\\
({\partial_t}-\eta D_1)B_R'&=&B_z^0 {\partial_z u_R'},\nn\\
({\partial_t}-\eta D_1 )B_{\phi}'&=&B_z^0{\partial_z u_{\phi}'}+\frac{2B_{\phi}^0}{R} u_R'+
(R{\partial_R \Omega})B_R',\nn\\
({\partial_t}-\eta D_2)B_z'&=&
-B_z^0 \partial_R^{\dagger} u_R',\nn\\
{\partial_z u_z'}&=&-\partial_R^{\dagger} u_R',\nn\\
{\partial_z B_z'}&=&-\partial_R^{\dagger} B_R',
\ea
where the squared epicyclic frequency $\kappa$ is defined as
\be{al2}
\kappa^2=2\Omega
\left(2\Omega +R \frac{d\Omega}{dR} \right)=\frac{1}{R^3}\frac{d}{dR}(\Omega^2 R^4).
\ee

Following the approach of \cite{G02,LGHJ06} we act on the first of
equations \rf{al1} by the operator $\partial_R^{\dagger}$
and on the third one by
the operator $\partial_z$.
Summing the results, taking into account that
\be{al3}
\partial_R^{\dagger}D_1=D_2 \partial_R^{\dagger},\quad \partial_R^{\dagger}B_{\phi}^0=0,\quad
\partial_R(B_{\phi}^0B_{\phi}')=-\frac{B_{\phi}^0B_{\phi}'}{R}+
B_{\phi}^0\partial_RB_{\phi}',
\ee
and using the last two equations of \rf{al1} yields
\ba{al4}
-2\Omega \partial_R^{\dagger} u_{\phi}'&=&-\frac{1}{\rho}D_2\left[p'+\frac{1}{\mu_0}B_z^0B_z'\right]-
\frac{1}{\rho}\frac{1}{\mu_0}(\partial_R^{\dagger}B_{\phi}^0\partial_R +\partial_z^2B_{\phi}^0)B_{\phi}'-
\frac{1}{\mu_0 \rho} \partial_R^{\dagger}\left( \frac{B_{\phi}^0}{R} B_{\phi}'\right)\nn\\
&=&-\frac{1}{\rho}D_2\left[p'+\frac{1}{\mu_0}(B_z^0B_z'+B_{\phi}^0B_{\phi}')\right]-
\frac{2}{\mu_0 \rho} \partial_R^{\dagger}\left( \frac{B_{\phi}^0}{R} B_{\phi}'\right).
\ea
Therefore, we extend the identity obtained in \cite{G02} to the case $B_{\phi}^0\ne0$
\be{al5}
D_2\frac{1}{\rho}\left[p'+\frac{1}{\mu_0}(B_z^0B_z'+B_{\phi}^0B_{\phi}')\right]=2 \partial_R^{\dagger} \left(\Omega u_{\phi}'
-
\frac{1}{\mu_0 \rho} \frac{B_{\phi}^0}{R} B_{\phi}'\right).
\ee

On the other hand, using \rf{al3} we transform
the first of the equations \rf{al1} into
\ba{al6}
(\partial_t-\nu D_1) u_R'-2\Omega u_{\phi}'
&=&
-\partial_R\frac{1}{\rho}\left[ p'+\frac{1}{\mu_0}\left(B_z^0B_z' + B_{\phi}^0B_{\phi}'\right)\right]\nn\\
&+&
\frac{1}{\mu_0 \rho} \left( B_z^0 {\partial_z}B_R' -\frac{2B_{\phi}^0}{R} B_{\phi}'\right).
\ea
Acting on both sides of the equation \rf{al6} by the operator $D_1$ and taking into account the identity \rf{al5} and
\be{al7}
D_1\partial_R=(\partial_R\partial_R^{\dagger}+\partial_z^2)\partial_R
=\partial_R(\partial_R^{\dagger}\partial_R+\partial_z^2)=\partial_R D_2
\ee
we get
\ba{al8}
D_1(\partial_t-\nu D_1) u_R'&-&2\Omega D_1u_{\phi}'\nn\\
&=&-\partial_R D_2\frac{1}{\rho}\left[ p'+\frac{1}{\mu_0}\left(B_z^0B_z' + B_{\phi}^0B_{\phi}'\right)\right]
+
D_1\frac{1}{\mu_0 \rho} \left( B_z^0 {\partial_z}B_R' -\frac{2B_{\phi}^0}{R} B_{\phi}'\right)\nn\\
&=&-2\partial_R  \partial_R^{\dagger} \left(\Omega u_{\phi}'
-
\frac{1}{\mu_0 \rho}  \frac{B_{\phi}^0}{R} B_{\phi}'\right)
+
D_1\frac{1}{\mu_0 \rho} \left( B_z^0 {\partial_z}B_R' -\frac{2B_{\phi}^0}{R} B_{\phi}'\right).
\ea
Rearranging the terms and using the definition of the operator $D_1$ yields
\ba{al9}
D_1(\partial_t-\nu D_1) u_R'-2\Omega \partial_z^2  u_{\phi}'
&=&2\partial_R  \partial_R^{\dagger}
\frac{1}{\mu_0 \rho}  \frac{B_{\phi}^0}{R} B_{\phi}'
+
 \frac{1}{\mu_0 \rho}B_z^0 D_1{\partial_z}B_R' -\frac{1}{\mu_0 \rho}D_1\frac{2B_{\phi}^0}{R} B_{\phi}'\nn\\
&=&
 \frac{1}{\mu_0 \rho}B_z^0 D_1{\partial_z}B_R' -\frac{1}{\mu_0 \rho}\frac{2B_{\phi}^0}{R} \partial_z^2B_{\phi}'.
\ea
Therefore, we have separated the equations for $u_R'$, $u_{\phi}'$ and $B_R'$, $B_{\phi}'$ from the others in \rf{al1}
\ba{al10}
(\partial_t-\nu D_1)D_1 u_R'-2\Omega \partial_z^2  u_{\phi}'
&=&
 \frac{1}{\mu_0 \rho}B_z^0 D_1{\partial_z}B_R' -\frac{1}{\mu_0 \rho}\frac{2B_{\phi}^0}{R} \partial_z^2B_{\phi}',\nn\\
({\partial_t}-\nu D_1)u_{\phi}'+\frac{\kappa^2}{2\Omega}u_R'&=&
\frac{B_z^0}{\mu_0\rho} {\partial_z B_{\phi}'},\nn\\
({\partial_t}-\eta D_1)B_R'&=&B_z^0 {\partial_z u_R'},\nn\\
({\partial_t}-\eta D_1 )B_{\phi}'&=&B_z^0{\partial_z u_{\phi}'}+\frac{2B_{\phi}^0}{R} u_R'+
(R{\partial_R \Omega})B_R'.
\ea
Note that after introducing the stream functions for the poloidal components
\be{al11}
u_R'=\partial_z\varphi,\quad u_z'=-\partial_R^{\dagger}\varphi,\quad B_R'=\partial_z \psi,\quad B_z'=-\partial_R^{\dagger}\psi,
\ee
the equations \rf{al10} extend the inviscid equations of \cite{LGHJ06} to the case $\nu\ne0$.

We can rewrite \rf{al10} in the form of the operator matrix equation
$\partial_t \tilde{E} \xi'=\tilde{H}\xi'$, where                   
$\xi'=(u_R',u_{\phi}',B_R',B_{\phi}')^T$,
\be{al12}
\tilde{E}=\left(                                    
    \begin{array}{cccc}
      D_1 & 0 & 0 & 0 \\
      0 & 1 & 0 & 0 \\
      0 & 0 & 1 & 0 \\
      0 & 0 & 0 & 1 \\
    \end{array}
  \right), \quad
\tilde{H}=\left(                               
    \begin{array}{cccc}
      \nu D_1^2 & 2\Omega \partial_z^2 & \frac{B_z^0}{\mu_0 \rho} D_1{\partial_z} & -\frac{2B_{\phi}^0}{\mu_0 \rho R} \partial_z^2 \\
      -\frac{\kappa^2}{2\Omega} & \nu D_1 & 0 & \frac{B_z^0}{\mu_0\rho} {\partial_z} \\
      B_z^0 \partial_z & 0 & \eta D_1 & 0 \\
      \frac{2B_{\phi}^0}{R} & B_z^0 \partial_z & R{\partial_R \Omega} & \eta D_1 \\
    \end{array}
  \right).
\ee
The resulting multiparameter family of operator matrices equipped with boundary conditions can be investigated by numerical
or perturbative \cite{Ki09a} methods. In the following we use the local WKB approximation.

\subsection{Local WKB approximation}

We choose a fiducial point $(R_0,z_0)$, around which we perform the local stability analysis \cite{PP05}.
We expand all the background quantities in Taylor series around $(R_0,z_0)$ and retain only
the zeroth order in terms of the local coordinates $\tilde R=R-R_0$ and $\tilde z=z-z_0$
to obtain the operator matrix equation with the constant coefficients
\ba{w0}
\partial_t \tilde{E}_0 \xi'=\tilde{H}_0\xi'                       
\ea
with
\be{w1}
\tilde{E}_0=\left(                                                
    \begin{array}{cccc}
      D_1^0 & 0 & 0 & 0 \\
      0 & 1 & 0 & 0 \\
      0 & 0 & 1 & 0 \\
      0 & 0 & 0 & 1 \\
    \end{array}
  \right), \quad
\tilde{H}_0=\left(                                           
    \begin{array}{cccc}
      \nu ({D_1^0})^2 & 2\Omega_0 \partial_{\tilde z}^2 & \frac{B_z^0}{\mu_0 \rho} D_1^0{\partial_{\tilde z}} &
      -\frac{2B_{\phi}^0}{\mu_0 \rho R_0} \partial_{\tilde z}^2 \\
      -\frac{\kappa_0^2}{2\Omega_0} & \nu D_1^0 & 0 & \frac{B_z^0}{\mu_0\rho} {\partial_{\tilde z}} \\
      B_z^0 \partial_{\tilde z} & 0 & \eta D_1^0 & 0 \\
      \frac{2B_{\phi}^0}{R_0} & B_z^0 \partial_{\tilde z} & \frac{\kappa_0^2}{2\Omega_0}-2\Omega_0 & \eta D_1^0 \\
    \end{array}
  \right),
\ee
where
\be{w2}
\Omega_0=\Omega(R_0),\quad \kappa_0^2=2\Omega_0
\left(2\Omega_0 +R_0 \left.\frac{d\Omega}{dR}\right|_{R=R_0} \right),\quad B_{\phi}^0=B_{\phi}^0(R_0), \quad
D_1^0=\partial_{\tilde R}^2+\partial_{\tilde z}^2+\frac{\partial_{\tilde R}}{R_0}-\frac{1}{R_0^2}.
\ee

Equation \rf{w0} is a linear PDE with the constant coefficients in the local variables $(\tilde R, \tilde z)$
for the perturbed quantities $\xi '$. This is a good approximation as long as the variations $\tilde R$ and $\tilde z$ are
small in comparison with the characteristic length scales in the radial and vertical directions, respectively \cite{PP05}.
A solution to the equation \rf{w0} has the form of a plane wave            
\ba{w3}
{\xi}'&=&\tilde{\xi}\exp{(\gamma t + i k_R \tilde R + i k_z \tilde z)},\quad \tilde
\xi=(\tilde u_R,\tilde u_{\phi},\tilde B_R,\tilde B_{\phi})^T,
\ea
where $\tilde{\xi}$ is a vector of constant coefficients.

Introducing the total wave number $k^2=k_z^2+k_R^2$ and denoting $\alpha=k_z/k$, we find
\ba{w4}
 D_1^0 \exp{(\gamma t + i k_R \tilde R + i k_z \tilde z)} &=&\left(-k^2+\frac{ik_R}{R_0}-\frac{1}{R_0^2}\right)
 \exp{(\gamma t + i k_R \tilde R + i k_z \tilde z)}.
\ea
In the WKB approximation we restrict the analysis to the modes with the wave numbers satisfying $k_R R_0 \gg 1$ which allows us to
neglect the terms $\frac{ik_R}{R_0}-\frac{1}{R_0^2}$ in \rf{w4}. In view of this, after substitution of \rf{w3} into
equation \rf{w0}, we arrive at the
matrix eigenvalue problem
\be{w5}
(H-\gamma E)\tilde \xi=0,
\ee
with $E$ as a unit matrix and $H=-{\rm diag}(\omega_{\nu},\omega_{\nu},\omega_{\eta},\omega_{\eta})+H_1+H_2$, where
$\omega_{\nu}=\nu k^2$ and $\omega_{\eta}=\eta k^2$ are the viscous and resistive frequencies,
\be{w6}
H_1=\frac{i\omega_{A}}{\sqrt{\mu_0 \rho}}\left(
      \begin{array}{cccc}
        0 & 0 & 1 & 0 \\
        0 & 0 & 0 & 1 \\
        \mu_0 \rho & 0 & 0 & 0 \\
        0 & \mu_0 \rho & 0 & 0 \\
      \end{array}
    \right),
\ee
\be{w6a}
H_2=\left(
      \begin{array}{cccc}
        0 & 2\Omega_0 \alpha^2 & 0 & -2\omega_{A_{\phi}}\frac{\alpha^2}{\sqrt{\mu_0\rho}} \\
        -2\Omega_0 -R_0 \left.\frac{d\Omega}{dR}\right|_{R=R_0} & 0 & 0 & 0 \\
        0 & 0 & 0 & 0 \\
        -2\omega_{A_{\phi}}{\sqrt{\mu_0\rho}} & 0 & R_0 \left.\frac{d\Omega}{dR}\right|_{R=R_0} & 0 \\
      \end{array}
    \right),
\ee
and the Alfv\'en frequencies are
\be{w6b}
\omega_A^2=\frac{k_z^2 (B_z^0)^2}{\mu_0 \rho},\quad
\omega_{A_{\phi}}^2=\frac{(B_{\phi}^0)^2}{\mu_0\rho R_0^2}.
\ee

Note that the matrix $-{\rm diag}(\omega_{\nu},\omega_{\nu},\omega_{\eta},\omega_{\eta})+H_1$ has two
double eigenvalues related to the damped Alfv\'en modes  \cite{N09}
\be{w6c}
\gamma_{1,2}=-\frac{\omega_{\nu}+\omega_{\eta}}{2}+\sqrt{\left(\frac{\omega_{\nu}-\omega_{\eta}}{2} \right)^2-\omega_A^2},\quad
\gamma_{3,4}=-\frac{\omega_{\nu}+\omega_{\eta}}{2}-\sqrt{\left(\frac{\omega_{\nu}-\omega_{\eta}}{2} \right)^2-\omega_A^2}.
\ee

When $\omega_{A_{\phi}}=0$,
$\left.\frac{d\Omega}{dR}\right|_{R=R_0}=0$, the eigenvalues of the
matrix $H_1+H_2$ correspond to the Alfv\'en-inertial or
Magneto-Coriolis waves \cite{Le54} \be{me1}
\gamma_{1,2}=i\sqrt{\omega_A^2+\Omega_0^2\alpha^2}\pm
i\alpha\Omega_0,\quad
\gamma_{3,4}=-i\sqrt{\omega_A^2+\Omega_0^2\alpha^2}\pm
i\alpha\Omega_0. \ee Figure \ref{fig0} demonstrates how rotation leads        
to the splitting of plane Alfv\'en waves into the fast and slow
Magneto-Coriolis waves \cite{Le54}. The system with purely imaginary           
eigenvalues \rf{me1} is marginally stable and its destabilization
caused by dissipative, shear, and azimuthal magnetic field
perturbations admits thus a natural interpretation as a
dissipation-induced instability \cite{KM07,Ki07, Ki09}.

On the other hand, the matrix $H$ can be considered as a result of a
non-Hermitian complex perturbation $H_1+H_2$ of a real symmetric
matrix, which has two double semi-simple eigenvalues---diabolical
points \cite{BD03}. This is a typical situation for the problems of
wave propagation in chiral absorptive media \cite{KKM03,BD03,KMS05}
or in rotating symmetric continua \cite{Ki09}.

\subsection{Dispersion relation in terms of dimensionless parameters}

The stability of the propagating plane wave perturbation \rf{w4} is determined by the roots $\gamma$ of the dispersion relation
$P(\gamma)=\det(H-\gamma E)=0$, where
\be{d1}
P(\gamma)=\gamma^4+a_1\gamma^3+a_2\gamma^2+(a_3+ib_3)\gamma+a_4+ib_4=0.
\ee
We write the coefficients of the complex polynomial \rf{d1} in the form
\ba{d2}
a_1&=&2(\omega_{\nu}+\omega_{\eta}),\nn\\
a_2&=&(\omega_{\nu}+\omega_{\eta})^2+2(\omega_A^2+\omega_{\nu}\omega_{\eta})+\alpha^2\kappa_0^2+4\alpha^2 \omega_{A_{\phi}}^2,\nn\\
a_3&=&2(\omega_{\eta}+\omega_{\nu})(\omega_A^2+\omega_{\eta}\omega_{\nu})+
2\alpha^2\kappa_0^2\omega_{\eta}+4\alpha^2(\omega_{\eta}+\omega_{\nu})
\omega^2_{A_{\phi}},\nn\\
a_4&=&(\omega_A^2+\omega_{\nu}\omega_{\eta})^2-4\alpha^2\omega_A^2\Omega_0^2+\alpha^2\kappa_0^2(\omega_A^2+\omega_{\eta}^2)+
4\alpha^2\omega_{\nu}\omega_{\eta}\omega_{A_{\phi}}^2,\nn\\
b_3&=&-8\alpha^2\Omega_0\omega_A\omega_{A_{\phi}},\nn\\
b_4&=&-4\alpha^2\Omega_0\omega_A\omega_{A_{\phi}}(2\omega_{\eta}+\omega_{\nu})-\kappa_0^2\alpha^2 \Omega_0^{-1}\omega_A\omega_{A_{\phi}}(\omega_{\eta}-\omega_{\nu}).
\ea

After scaling the spectral parameter as $\gamma=\lambda \sqrt{\omega_{\nu}\omega_{\eta}}$, we express the appropriately normalized coefficients \rf{d2} by means of the dimensionless Rossby number $({\rm Ro})$, magnetic Prandtl number $({\rm Pm})$, ratio of the Alfv\'en frequencies $(\beta^*)$, Hartmann $({\rm Ha}^*)$, and Reynolds $({\rm Re}^*)$ numbers
\be{d3}
{\rm Ro}=\frac{1}{2}\frac{R_0}{\Omega_0}\left.\frac{d\Omega}{d R}\right|_{R=R_0},\quad
{\rm Pm}=\frac{\nu}{\eta}=\frac{\omega_{\nu}}{\omega_{\eta}},\quad
\beta^*=\alpha \frac{\omega_{A_{\phi}}}{\omega_A},\quad
{\rm Re^*}=\alpha\frac{\Omega_0}{\omega_{\nu}},\quad
{{\rm Ha}^*}=\alpha\frac{B_z^0}{k\sqrt{\mu_0 \rho \nu \eta}}.
\ee
Additional transformation yields the coefficients of the dispersion relation $P(\lambda)=0$ in a simplified form
\ba{d4}
a_1&=&2\left(\sqrt{\rm Pm} +\frac{1}{\sqrt{\rm Pm}} \right),\nn \\
a_2&=&\frac{a_1^2}{4}+2({1+{\rm Ha}^*}^2)+4{\beta^*}^2{{\rm Ha}^*}^2+4{\rm Re^*}^2   {\rm Pm} (1+{\rm Ro}),\nn \\
a_3&=&a_1(1+{{\rm Ha}^*}^2)+ 2 a_1 {\beta^*}^2{{\rm Ha}^*}^2+8{{\rm Re}^*}^2(1+{\rm Ro})\sqrt{\rm Pm},\nn \\
a_4
&=&\left(1+{{\rm Ha}^*}^2\right)^2+4{\beta^*}^2{{\rm Ha}^*}^2+4{\rm Re^*}^2+4{\rm Re^*}^2{\rm Ro}({\rm Pm}{{\rm Ha}^*}^2+1),\nn \\
b_3&=&-8\beta^*{{\rm Ha}^*}^2{\rm Re}^*\sqrt{\rm Pm},\nn\\
b_4
&=&-4\beta^*{{\rm Ha}^*}^2{\rm Re}^*(2+(1-{\rm Pm}){\rm Ro}).
\ea
Therefore, we have exactly reproduced the dispersion relation of \cite{LV07,RS08},
which generalizes that of \cite{G02,LGHJ06}.

\section{SMRI in the absence of the azimuthal magnetic field $(\beta^*=0)$}

Let us first assume $\beta^*=0$ and study the onset of the standard magnetorotational instability (SMRI).
The coefficients of the polynomial $P(\lambda)$ are then real because $b_3=0$ and $b_4=0$.
We have
\ba{s1}
a_1&=&\hat a_1=2\left(\sqrt{\rm Pm} +\frac{1}{\sqrt{\rm Pm}} \right),\nn \\
a_2&=&\hat a_2=\frac{\hat a_1^2}{4}+2({1+{\rm Ha}^*}^2)+4{\rm Re^*}^2   {\rm Pm} (1+{\rm Ro}),\nn \\
a_3&=&\hat a_3=\hat a_1(1+{{\rm Ha}^*}^2)+8{{\rm Re}^*}^2(1+{\rm Ro})\sqrt{\rm Pm},\nn \\
a_4&=&\hat a_4
=\left(1+{{\rm Ha}^*}^2\right)^2+4{\rm Re^*}^2(1+{\rm Ro}({\rm Pm}{{\rm Ha}^*}^2+1)).
\ea

Composing the Hurwitz matrix of the real polynomial $P(\lambda)$ we write the Lienard and Chipart criterion of asymptotic stability \cite{LS14,M66}: all roots $\lambda$ have ${\rm Re}\lambda<0$ if and only if
\be{s2}
\hat a_4>0,\quad \hat a_2>0,\quad h_1=\hat a_1>0,\quad h_3=\hat a_1\hat a_1\hat a_3-\hat a_1^2\hat a_4-\hat a_3^2>0.
\ee
Explicit calculation of $h_3$ shows that it is a sum of squared quantities
\be{s3}
h_3=64\left({\rm Pm^*}{\rm Re^*}^2({\rm Ro}+1)+\frac{\hat a_1^2}{16}\right)^2+{\rm Ha^*}^2\hat a_1^2\left(\frac{\hat a_1^2}{4}+4{\rm Re^*}^2\right)>0
\ee
Therefore, the condition $h_3>0$ is always fulfilled.

The local definition of the Rossby number \rf{d3} allows us to vary it for the background profile $\Omega(R)=a+b R^{-2}$
changing the coefficients $a$ and $b$ because ${\rm Ro}=-b/(aR_0^2+b)$. On the other hand we can interpret the Rossby numbers
as if they would correspond to quite general rotation profiles $\Omega(R)$, which can have, e.g., the
shape $\Omega(R)\sim R^{w}$ (with $w=-3/2$ and ${\rm Ro}=w/2=-3/4$ for Kepler
rotation).

In the following we assume ${\rm Ro}\ge-1$ that corresponds to the centrifugally (Rayleigh) stable flow in the absence of the magnetic field. This reduces the conditions \rf{s2} to  $\hat a_4>0$ that is equivalent to
\be{s4}
{\rm Ro}>{\rm Ro}^c=-\frac{\left(1+{\rm Ha^*}^2\right)^2+4{\rm Re^*}^2}{4{\rm Re^*}^2({\rm Pm}{\rm Ha^*}^2+1)}.
\ee

Note that in the absence of the magnetic field, ${\rm Ha^*}=0$, the inequality \rf{s4} is        
\be{s5}
{\rm Ro}>{\rm Ro}^v=-1-\frac{1}{4{\rm Re^*}^2},
\ee
where we define the viscous Rayleigh line ${\rm Ro}^v$.
In the inviscid limit ${\rm Re}^*\rightarrow\infty$ it is reduced to the Rayleigh's centrifugal stability criterion
\be{s6}
{\rm Ro}>{\rm Ro}^i=-1,
\ee
where ${\rm Ro}^i$ is the classical inviscid Rayleigh line.

As is seen in the Figure \ref{fig1}(left), there are two extrema of the function ${\rm Ro}^c({\rm Ha}^*)$ at
\be{s7}
{\rm Ha^*}_{\max}=\pm \sqrt{\frac{-1+\sqrt{(1-{\rm Pm})^2+4{\rm Pm}^2{\rm Re^*}^2}}{\rm Pm}},
\ee
which agrees with the results of \cite{JGK01}. Triggered by                   
the vertical magnetic field $(B_z^0\ne 0)$ at some values
of ${\rm Ha}^*$ the flow becomes unstable for ${\rm Ro}>-1$.
It can be stabilized again, however, with the further increase of ${\rm Ha}^*$, which is a hallmark of the standard MRI, cf. with Figure 1 in \cite{JGK01}.

The maximal values of the Rossby number at the peaks of the boundary of the SMRI domain are
\be{s8}
{\rm Ro}_{\max}^c=-\frac{4{\rm Re^*}^2{\rm Pm}^2+\left({\rm Pm}-1+\sqrt{(1-{\rm Pm})^2+4{\rm Re^*}^2{\rm Pm}^2}\right)^2}{4{\rm Re^*}^2{\rm Pm}^2\sqrt{(1-{\rm Pm})^2+4{\rm Re^*}^2{\rm Pm}^2}}.
\ee
The two extrema at ${\rm Ha^*}_{\max}\ne 0$ exist when the radicand in \rf{s7} is positive, that is when
\be{s9}
{\rm Re^*}{\rm Pm}>\frac{\sqrt{(2-{\rm Pm}){\rm Pm}}}{2}.
\ee
Otherwise, the unique maximum is at the origin, Figure \ref{fig1}(right).
This condition also follows from the positiveness of the second-order coefficient in the series expansion of ${\rm Ro}^c({\rm Ha^*})$ at ${\rm Ha^*}=0$:
\be{s10}
{\rm Ro}^c=-\frac{1+4{\rm Re^*}^2}{4{\rm Re^*}^2}-\frac{2-(4{\rm Re^*}^2+1){\rm Pm}}{4{\rm Re^*}^2}{\rm Ha^*}^2+\cdots
\ee

Setting ${\rm Ro}_{\max}^c\ge-3/4$, we find the conditions for existence of the standard MRI at and above the Kepler line
in terms of the magnetic Reynolds number ${\rm Rm}^*={\rm Re^*Pm}$
\be{s11}
{\rm Rm}^*\ge\frac{2}{3}\sqrt{1+3{\rm Pm}}.
\ee
For ${\rm Pm}\ll 1$ one should have ${\rm Rm}^*\ge 2/3$ to have SMRI for the Kepler flows, which leads to ${\rm Re^*}\gg 1$.
At such values of $\rm Re^*$ the following formal asymptotic expansions of ${\rm Ha^*}_{\max}({\rm Re^*})$ and ${\rm Ro}_{\max}^c({\rm Re^*})$ are valid
\be{s12}
{\rm Ha^*}_{\max}=\pm\sqrt{2 \rm Re^*}\mp\frac{\sqrt{2}}{4{\rm Pm}\sqrt{\rm Re^*}}, \quad
{\rm Ro}_{\max}^c=-\frac{1}{\rm PmRe^*}+\frac{1-{\rm Pm}}{2{\rm Pm}^2{\rm Re^*}^2}-\frac{(1-{\rm Pm})^2}{2^3{\rm Pm}^3{\rm Re^*}^3}.
\ee
The asymptotic expansion \rf{s12} gives a simple scaling law                    
which is known to be a characteristic of SMRI
\be{s13}
{\rm Re^*}=\frac{1}{2}{\rm Ha^*}^2.
\ee
This equation is identical to
$\rm N^*:={\rm Ha^*}^2/{\rm Re^*}=2$
where $\rm N^*$ is often called
interaction parameter (in technical magnetohydrodynamics)
or Elsasser number (in geo- and astrophysics).

\begin{figure}
\includegraphics[width=160mm]{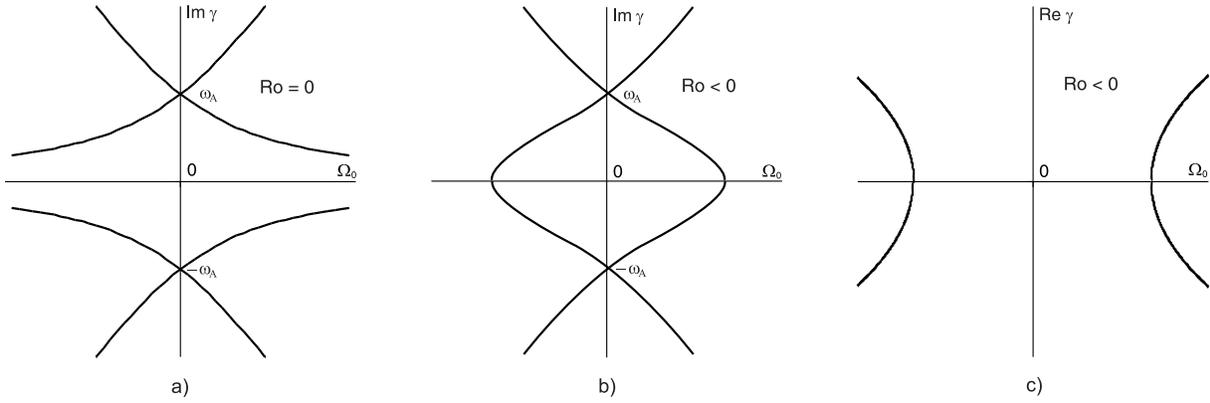}
\caption{\label{fig0} (a) Coriolis splitting of the Alfv\'en plane wave into the fast and the slow   
Magneto-Coriolis (MC) waves, (b) shear causes interaction of slow MC branches with the origination of
the double zero eigenvalue, (c) splitting of the double zero eigenvalue yields positive real eigenvalues (SMRI). }
\end{figure}

\begin{figure}
\includegraphics[width=160mm]{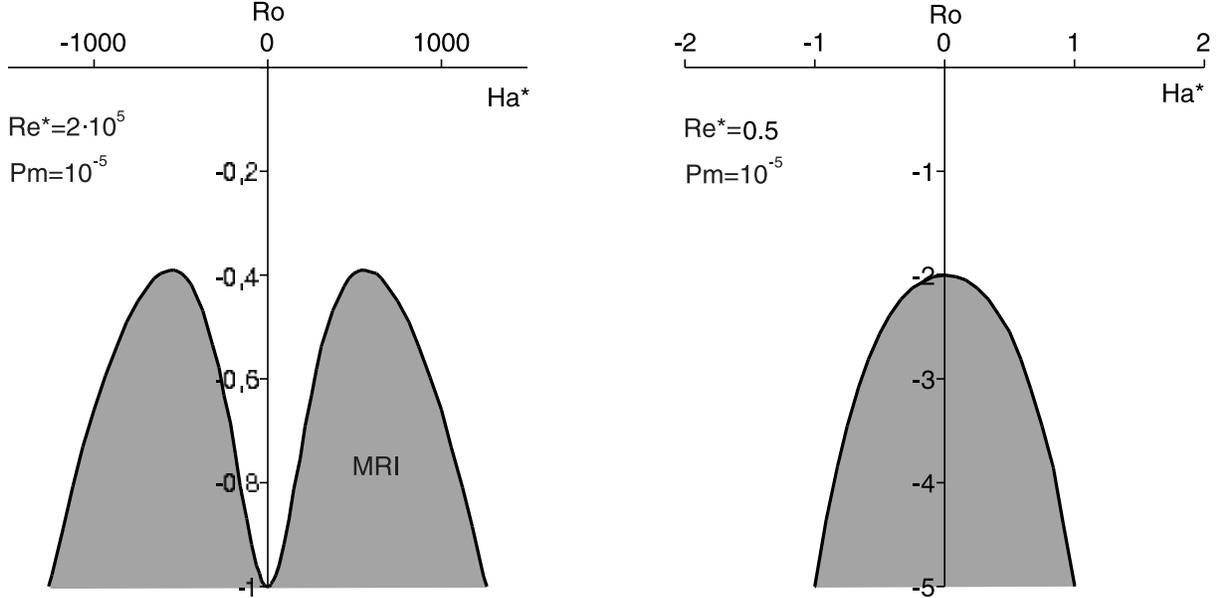}
\caption{\label{fig1} The domain according to equation                     
\rf{s4} of the standard MRI (grey) for ${\rm Pm}=10^{-5}$ and (left)
${\rm Re}^*=2\cdot10^5$ or (right) ${\rm Re}^*=0.5$. The left plot shows the typical SMRI peaks \cite{JGK01}.}
\end{figure}

The SMRI can be interpreted as destabilization of slow                           
Magneto-Coriolis waves \cite{N08,N09}. Indeed in the
presence of shear, ${\rm Ro}\ne0$, we find from the equation \rf{d1}
with the coefficients \rf{d2} that in the absence of dissipation
($\omega_{\nu}=0$, $\omega_{\eta}=0$) the eigenvalues are
\be{s14}
\gamma=\pm\sqrt{-2\Omega_0^2\alpha^2(1+{\rm Ro})-\omega_A^2\pm2\Omega_0\alpha\sqrt{\Omega_0^2\alpha^2(1+
{\rm Ro})^2+\omega_A^2}}.
\ee
At the critical value $\Omega_0=\Omega_0^c$, where
\be{s15}
\Omega_0^c=-\frac{\omega_A^2}{2\alpha^2R_0 \partial_R\Omega},
\ee
the branches of the slow magnetic-Coriolis
waves merge with the origination of the double zero
eigenvalue, see Figure \ref{fig0}(b).
Splitting of this eigenvalue yields positive
real eigenvalues, see Figure \ref{fig0}(c).                              
Note that the SMRI threshold \rf{s15} is equivalent to
\be{s16}
{\rm Ro}=-\frac{{\rm Ha^*}^2}{4{\rm Re^*}^2 {\rm Pm}}
\ee
that follows from  \rf{s4} when ${\rm Ha^*}\rightarrow \infty$.

\section{HMRI in the presence of an azimuthal                            
magnetic field $(\beta^* \ne 0)$}

The fact that an additional azimuthal field changes the
character of the MRI drastically had been detected by Knobloch
as early as 1992 \cite{K92}. He had shown
that in this case the instability appears in form of a
travelling wave (see also \cite{K96}). However, the
difference in the scaling behaviour for small $\rm Pm$
between standard and helical MRI was worked out only recently
\cite{HR05}, and is still the subject of intense debate.
In this section we will contribute to this discussion
by focusing on the specific $\rm Pm$ dependence of the helical MRI.         

\subsection{Bilharz criterion for asymptotic stability}
With the appearance of the azimuthal magnetic
field $(\beta^* \ne 0)$, the coefficients of the
polynomial $P(\lambda)$ become complex. This breaks
the symmetry of the eigenvalues with respect to the
real axis of the complex plane and consequently may lead to
dramatic changes in the stability properties of the system.

In contrast to previous studies \cite{LV07,RS08} that
were based on the study of approximations to the roots of the dispersion relation
\be{a0}
P(\lambda)=\lambda^4+a_1\lambda^3+a_2\lambda^2+(a_3+ib_3)\lambda+a_4+ib_4=0,
\ee
we prefer to use the Bilharz criterion \cite{Bi44,M66} of
asymptotic stability of the roots of complex polynomials.
This criterion establishes the necessary and sufficient
conditions for all the roots to be in the left part of
the complex plane $({\rm Re}\lambda < 0)$ in terms of
positiveness of the main even-ordered minors of the
Bilharz matrix. For the polynomial $P(\lambda)$ with the coefficients \rf{d4} this matrix is
\be{a1}
B=\left(
    \begin{array}{cccccccc}
       a_4 & -b_4 &  0   &  0   &  0   &  0   &  0   &  0 \\
       b_3 &  a_3 &  a_4 & -b_4 &  0   &  0   &  0   &  0 \\
      -a_2 &  0   &  b_3 &  a_3 &  a_4 & -b_4 &  0   &  0 \\
       0   & -a_1 & -a_2 &  0   &  b_3 &  a_3 &  a_4 & -b_4 \\
       1   &  0   &  0   & -a_1 & -a_2 &  0   &  b_3 &  a_3 \\
       0   &  0   &  1   &  0   &  0   & -a_1 & -a_2 &  0 \\
       0   &  0   &  0   &  0   &  1   &  0   &  0   & -a_1 \\
       0   &  0   &  0   &  0   &  0   &  0   &  1   &  0 \\
    \end{array}
  \right).
\ee

The Bilharz stability conditions \cite{Bi44,M66} require positiveness of all diagonal even-ordered minors of $B$
\ba{a2}
m_1&=&a_3a_4+b_3b_4>0, \\
m_2&=&(a_2a_3-a_1a_4)m_1-a_2^2b_4^2>0,\nn \\
m_3&=&(a_1a_2-a_3)m_2-(a_1^2a_4a_2+(a_1b_3-b_4)^2)m_1
+a_1a_4(b_4a_2(2b_4-a_1b_3)+a_1^2a_4^2)>0,\nn \\
m_4&=&a_1m_3-a_1a_3m_2+ (a_3^3+a_1^2b_4b_3-2a_1b_4^2)m_1+a_1b_4^2a_4(a_1a_2 -a_3)-b_4^2a_3^2a_2+b_4^4>0.\nn
\ea
The inequalities \rf{a2} determine the stability condition                
of the general dispersion relation \rf{a0}
in the presence of both vertical $(B_z^0)$ and azimuthal $(B_{\phi}^0)$ components of the magnetic field.

We first note that for $\beta^*=0$ the stability                       
conditions \rf{a2} are reduced to the stability condition $\hat a_4 > 0$ that was derived in the previous section. Indeed, with $\beta^*=0$ the coefficients $b_3$ and $b_4$ vanish to zero, which yields
\ba{a3}
m_1=\hat a_4\hat a_3,\quad
m_2=\hat a_4\hat a_3(\hat a_2\hat a_3-\hat a_1\hat a_4),\quad
m_3=\hat a_4(\hat a_2\hat a_3-\hat a_1\hat a_4)h_3,\quad
m_4=\hat a_4h_3^2.
\ea

In view of $\hat a_3>0$ and $h_3>0$ it remains to check the sign
of the expression $\hat a_2\hat a_3-\hat a_1\hat a_4$.
Explicit calculation yields
\ba{a4}
\frac{\hat a_2\hat a_3-\hat a_1\hat a_4}{2}&{=}& \frac{(4{\rm Pm}{\rm Re^*}^2({\rm Ro}+1)+{\rm Pm}+{\rm Ha^*}^2+2)^2{\rm Pm}}{{\rm Pm}\sqrt{\rm Pm}}\\
&{+}&  \frac{(4{\rm Re^*}^2{\rm Pm}^2+1)({\rm Pm}+1){\rm Ha^*}^2{+}{\rm Pm}^2{\rm Ha^*}^2({\rm Ha^*}^2+3+{\rm Pm})+1}{{\rm Pm}\sqrt{\rm Pm}}>0.\nn
\ea
Therefore, for $\beta^*=0$ the conditions \rf{a3} are            
reduced to the inequality $\hat a_4 > 0$ that determines
the stability domain that is adjacent to the domain of SMRI.

\begin{figure}
\includegraphics[width=80mm]{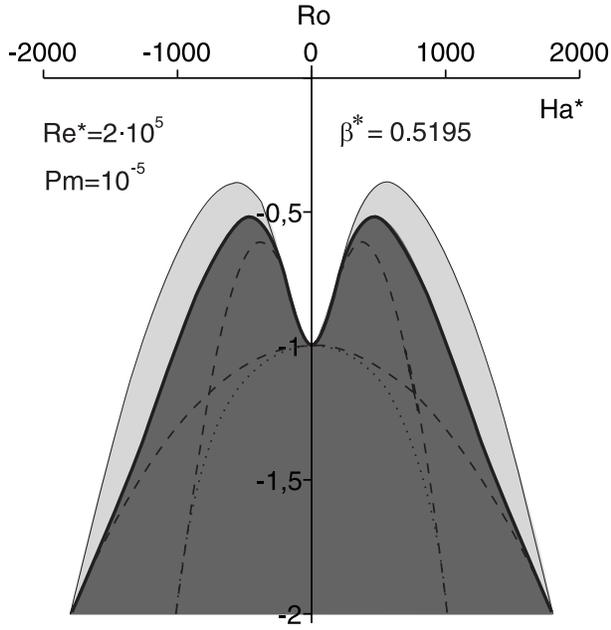}
\caption{\label{fig2} The domain according to equation                       
\rf{s4} of SMRI (light grey, $\beta^*=0$) and that of HMRI (dark grey)
for ${\rm Pm}=10^{-5}$ and ${\rm Re}^*=2\cdot10^5$. The thin full line marks    
the stability boundary ${\rm Ro}={\rm Ro}^c$.
The dashed, dotted, and bold lines bound the domains where the Bilharz
minors $m_{1,2,3,4}>0$.
The domain of HMRI which is adjacent to the intersection of these
domains is shown in dark grey.
The boundary to the HMRI domain (bold line) is $m_4=0$. }
\end{figure}

In Figure \ref{fig2} we plot the boundary \rf{s4} of            
SMRI domain to compare it with the domain of HMRI given by the
inequalities \rf{a2}. We see that the domain $m_4>0$ is an
intersection of all the domains $m_i>0$, $i=1,2,3,4$. Thus, the
region of HMRI, shown by dark grey in Figure \ref{fig2}, is adjacent
to the domain $m_4>0$. Although this fact is not a proof that the
inequalities \rf{a2} are reduced to the last one, our numerical
computations of the domains and of the roots of the dispersion
relation as well as the analysis of the inductionless approximation
in the next section confirm that $m_4=0$ is the boundary of HMRI
domain.

\subsection{Inductionless approximation}

As it was first observed in \cite{PGG07}, a remarkable feature of HMRI is    
that it
leads to destabilization, even in the
limit ${\rm Pm} \rightarrow 0$, for
some ${\rm Ro}>-1$, although not until the
Kepler profile $({\rm Ro}=-0.75)$. Below we prove this.

Let us consider the Rossby number as a function of the magnetic Prandtl number and fix all other parameters.
Substituting ${\rm Ro}=a+b{\rm Pm}+\ldots$ into the equation $m_4=0$ and collecting the terms with the identical
powers of $\rm Pm$, we find the quadratic equation on the coefficient $a$, which can be exactly solved. Therefore,
in the limit ${\rm Pm} \rightarrow +0$ there are two branches of the function ${\rm Ro}({\rm Ha^*}, {\rm Re^*}, {\rm \beta^*})$:
a positive one $({\rm Ro}^{+}>0)$ and a negative one $({\rm Ro}^{-}<0)$
\ba{i1}
& &{\rm Ro}^{\pm}\nn\\ &=& \frac{\left(1{+}{\rm Ha^*}^2\right)^2{+}4{\beta^*}^2{\rm Ha^*}^2(1{+}{\beta^*}^2 {\rm Ha^*}^2)}{2{\beta^*}^2{\rm Ha^*}^4}\\
&\pm&\frac{(2{\beta^*}^2 {\rm Ha^*}^2{+}{\rm Ha^*}^2{+}1)
\sqrt{\left(1{+}{\rm Ha^*}^2\right)^2{+}4{\beta^*}^2{\rm Ha^*}^2(1{+}{\beta^*}^2{\rm Ha^*}^2){+}
\frac{{\rm Ha^*}^4{\beta^*}^2}{{\rm Re^*}^2}\left(\left(1{+}{\rm Ha^*}^2\right)^2{+}4{\beta^*}^2{\rm Ha^*}^2\right)}}{2{\rm Ha^*}^4{\beta^*}^2}.\nn
\ea
When ${\beta^*}\rightarrow +0$ the function ${\rm Ro}^{+}$ tends to infinity while for ${\rm Ro}^-$ we get
\be{i2}
\lim_{{\beta^*}\rightarrow +0}{\rm Ro}^-=-1-\frac{\left(1+{\rm Ha^*}^2\right)^2}{4{\rm Re^*}^2}.
\ee
The expression \rf{i2} can also be obtained as a limit of ${\rm Ro}^c$ defined in \rf{s4} when ${\rm Pm} \rightarrow +0$.

Calculating the derivative $\frac{\partial{\rm Ro}^{-}}{\partial \rm Re^*}$ we find that it is strictly positive
for all ${\rm Re^*} \in (0, +\infty)$
\ba{i3}
& &\frac{\partial{\rm Ro}^{-}}{\partial \rm Re^*}\nn\\&=&\frac{(1+{\rm Ha^*}^2)^3+{\beta^*}^2{\rm Ha^*}^2(6+8{\rm Ha^*}^2+2{\rm Ha^*}^4+8{\beta^*}^2{\rm Ha^*}^2)}
{2{\rm Re^*}^2\sqrt{{\rm Re^*}^2((1+{\rm Ha^*}^2)^2+4{\beta^*}^2{\rm Ha^*}^2(1+{\beta^*}^2{\rm Ha^*}^2))+{\rm Ha^*}^4{\beta^*}^2((1+{\rm Ha^*}^2)^2+4{\beta^*}^2{\rm Ha^*}^2)}}\nn \\ &>&0.
\ea

Consequently, the maximal value of ${\rm Ro}^{-}<0$ is attained when ${\rm Re^*}\rightarrow +\infty$. In this limit
the function ${\rm Ro}^{-}(\beta^*)$ has a maximum
\be{i4}
{\rm Ro}^{-}_{\max}({\rm Ha^*})=
\max_{\beta^*}\lim_{{\rm Re^*}\rightarrow +\infty}{\rm Ro}^{-}=
-\frac{-{\rm Ha^*}^2-2+\sqrt{4+6{\rm Ha^*}^2+2{\rm Ha^*}^4}}{2{\rm Ha^*}^2}
\ee
at
\be{i5}
\beta^*_{\max}({\rm Ha^*})=\frac{\sqrt{2+2{\rm Ha^*}^2}}{2{\rm Ha^*}}.
\ee
Since the derivative of the function ${\rm Ro}^-_{\max}({\rm Ha^*})$ is strictly positive for all ${\rm Ha^*}>0$
\be{i6}
\frac{d {\rm Ro}^-_{\max}({\rm Ha^*})}{d {\rm Ha^*}}=
4\frac{3{\rm Ha^*}^2+4-2\sqrt{4+6{\rm Ha^*}^2+2{\rm Ha^*}^4}}{{\rm Ha^*}^3\sqrt{4+6{\rm Ha^*}^2+2{\rm Ha^*}^4}}>0
\ee
we conclude that the global maximum of the function ${\rm Ro}^-({\rm Ha^*},{\rm Re^*},\beta^*)$ coincides with the maximal
value of ${\rm Ro}^-_{\max}({\rm Ha^*})$, which is attained at ${\rm Ha^*}\rightarrow +\infty$ and is therefore
\be{i7}
\max_{{\rm Ha^*}, {\rm Re^*}, {\rm \beta^*}}{\rm Ro}^{-}=\max_{{\rm Ha^*}}{\rm Ro}^{-}_{\max}({\rm Ha^*})=
2-2\sqrt{2}\simeq-0.8284,
\ee
being exactly the same value that was found in the highly resistive inviscid limit in \cite{LGHJ06}.
The corresponding optimal value of $\beta^*$ in the limit ${\rm Ha^*}\rightarrow +\infty$ is
\be{i8}
\beta^*_{\max}=\frac{\sqrt{2}}{2}\simeq0.7071.
\ee
Note that numerical maximization of $\rm Ro$ frequently leads to the extrema corresponding to the values of $\beta^*\simeq 0.7$
even for $\rm Pm \ne 0$.

Extending the inviscid result of \cite{LGHJ06} we establish that in the inductionless approximation $(\rm Pm=0)$
the upper bound for HMRI is
\be{i9}
{\rm Ro}^{-}({\rm Ha^*}, {\rm Re^*}, {\rm \beta^*})<2-2\sqrt{2}\simeq-0.8284.
\ee
Proceeding similarly, we find that
\be{i10}
{\rm Ro}^{+}({\rm Ha^*}, {\rm Re^*}, {\rm \beta^*})>2+\sqrt{2}\simeq4.8284,
\ee
where the minimum is attained at the same extremal value of $\beta^*$ given by \rf{i8}.
The lower bound \rf{i10} for ${\rm Ro}^{+}$ exactly coincides with that found in \cite{LGHJ06}
in the highly resistive inviscid limit by analyzing the roots of the dispersion relation.
However, it should be noticed that the character of this ${\rm Ro}^{+}$ is still
unclear. Since up to present we have not obtained any corresponding result
from a 1D global eigenvalue solver, it remains to be checked if
this result is an artefact of the short wavelength approximation.

Anyway, quite in accordance with \cite{LV07,LGJ07}
we conclude that in the inductionless approximation there is no HMRI for
\be{i11}
2-2\sqrt{2}<{\rm Ro}({\rm Ha^*}, {\rm Re^*}, {\rm \beta^*})<2+2\sqrt{2},
\ee
which excludes HMRI for the Kepler law and for
other shallower velocity profiles.

Finally, we would like to find a
scaling law for HMRI to compare it with that of SMRI \rf{s13}.
The HMRI scaling law for the maximum of the critical
Rossby number at infinity (which works well, however,
starting from
${\rm Ha^*}\simeq3$) reads                               
\be{i5}
{\rm Re}^*=\left(1+\frac{\sqrt{2}}{2}\right){\rm Ha^*}^3.
\ee
In terms of the interaction parameter, this can be rewritten
as $\rm N^* Ha^*=1/\left(1+\sqrt{2}/2\right)$. This scaling is
rather different from the scaling of SMRI (Eq. 45).

\subsection{HMRI in the case when ${\rm Pm}\ne0$}                

In the previous section we have confirmed that for $\rm Pm=0$,
HMRI does not work for Keplerian flows, at
least according to the WKB approximation. Nevertheless, Hollerbach        
and R\"udiger had shown that it does
when considered as an eigenvalue problem, provided that at least one         
of the
radial boundary is conducting \cite{RH07}.

In this section we analyze the dispersion relation without
the simplifying assumption that $\rm Pm = 0$. As it follows from the
equation \rf{s4}, in the absence of the azimuthal component
of the magnetic field ($\beta^*=0$) there is no SMRI
for ${\rm Pm}<{\rm Pm}^c$, where at the threshold
\be{mp0}
{\rm Pm}={\rm Pm}^c:=- \frac{\left(1+{{\rm Ha}^*}^2\right)^2+4{{\rm Re}^*}^2(1+{\rm Ro})}{4{\rm Ro}{{\rm Re}^*}^2{{\rm Ha}^*}^2}.
\ee
The SMRI in this case develops when ${\rm Pm}>{\rm Pm}^c$.
In Figure \ref{fig9} the threshold \rf{mp0} is shown by the dashed line.

\subsubsection{First scenario of HMRI excitation}

When the azimuthal magnetic field is switched on ($\beta^* \ne 0$),
the instability threshold ${\rm Pm}^c$ depends on $\beta^*$.
At small values of $\beta^*$ the threshold slightly increases so that
the boundary of the domain of instability bends to the right of the
dashed line in the $({\rm Pm},\beta^*)$--plane, see Figure \ref{fig9}.    
The behavior of the instability boundary with  further increase          
of $\beta^*$ is determined by the Rossby number.

For Rossby numbers close to ${\rm Ro}=-1$ the instability
boundary at some $\beta^* \ne 0$ bends
to the left, crosses the dashed line and forms a ''semi-island''
of instability with its center
located close to $\beta^*=\frac{\sqrt{2}}{2}$, Figure \ref{fig9}(a).        
This enlargement of the instability domain on the left of
the dashed line is entirely a consequence of significant
helicity of the magnetic field. Although the whole grey area
of instability in Figure \ref{fig9}(a) is of course the
region of the helical magneto-rotational instability
(HMRI), the main interesting effect of non-zero helicity
is the excitation of instability at small and even infinitesimally
small values of ${\rm Pm}$ in the range where SMRI is not possible.    
With the increase of $\rm Ro$ the specific effect of
$\beta^*$ becomes even more pronounced with the bifurcation of
the instability domain to the isolated region (island) that lies
on the left of the dashed line and to the "continent" on the right
of it, Figure \ref{fig9}(b,c). The island of HMRI illustrates both
the destabilizing role of the azimuthal magnetic field component
and the non-triviality of inducing HMRI below the threshold \rf{mp0}         
at small negative values of the Rossby number $({\rm Ro}>-1)$. For
these reasons, we propose to refer to the instability for
$\beta^*\ne 0$ on the left of the dashed line as the
{\it essential HMRI} and that on the right as the
{\it helically modified SMRI}.

Despite the apparent discontinuities in the  $(\beta^*, \rm                  
Ro)$-plane, the three-dimensional domain of HMRI in the $(\beta^*,
{\rm Pm}, \rm Ro)$-space has a smooth boundary given by the
expression $m_4=0$, Figure \ref{fig14}(a). As it is seen in Figure
\ref{fig14}(a), the function ${\rm Ro}^c({\rm Pm},\beta^*)$ has
local extrema at some ${\rm Pm}\ne 0$, yielding regions of HMRI that
are separated from each other in the plane $({\rm Pm},\beta^*)$,
Figure \ref{fig9}(b,c). Since the maximum is attained at small but
finite values of $\rm Pm$, the corresponding boundary of HMRI in the
$(\beta^*, \rm Ro)$-plane at ${\rm Pm}\ne 0$ can exceed that in the
inductionless limit (an instability induced by the viscosity
$\omega_{\nu}\ne0$) and, moreover, the limiting bound $\rm Ro =
2-2\sqrt{2}$, as is clearly seen in Figure \ref{fig14}. The
one-dimensional slices in the $(\beta^*, \rm Ro)$-plane converge
however to the region of the inductionless HMRI when $\rm Pm
\rightarrow 0$. Therefore, in comparison to
 the inductionless limit, for ${\rm Pm}\ne 0$ we obtain higher values of the maximal Rossby numbers
 corresponding to the excitation of HMRI---a quite promising similarity  of this local WKB         
analysis with the
 observation of HMRI for Keplerian flows with conducting boundaries
in \cite{RH07}. In Figure \ref{fig15}, the evolution
 of the stability boundaries in the $({\rm Pm}, {\rm Re}^*)$-plane with the increase of $\beta^*$ demonstrates
 the details of the mechanism of reduction of the critical Reynolds number, which is another important characteristic
 of HMRI.

To clarify the nature of HMRI and SMRI and their
relation to each other we inspect now the roots of
the dispersion relation as functions of ${\rm Pm}$.                    
Series
expansions of the roots in the vicinity of ${\rm Pm}=0$ at $\beta^*=0$ yield
\ba{mp1}
\lambda_{1,3}&=&\left[-1-{\rm Ha^*}^2\pm2{\rm Re^*}\sqrt{-(1+{\rm Ro})}\right]\sqrt{\rm Pm}+o({\rm Pm^{1/2}}),\nn\\
\lambda_{2,4}&=&-\frac{1}{\sqrt{\rm Pm}}+{\rm Ha^*}^2\sqrt{\rm Pm}+o({\rm Pm^{1/2}})
\ea
Therefore, two eigenvalues $\lambda_{1,3}$ branch from zero and the other two $\lambda_{2,4}$ branch from minus infinity.
The eigenvalues $\lambda_{2,4}$ are real and negative, whereas the eigenvalues $\lambda_{1,3}$ are real in the vicinity
of the origin for ${\rm Ro}\le-1$ and complex otherwise with the frequency
$$
\omega=2\Omega_0 \frac{k_z}{k}\sqrt{{\rm Ro}+1}.
$$
The eigenvalues $\lambda_{1,3}$ correspond to inertial waves with $
\omega=\omega_{\rm C}:=2\Omega_0 {k_z}/{k}
$, if we assume rotation without shear $({\rm Ro}=0)$ and without damping, see e.g. \cite{N09}.

In the particular case, when $\beta^*=0$ and ${\rm Ro}=-1$ the dispersion equation is exactly solved
\ba{mp2}
\lambda_{1,2}&=&\frac{-{\rm Pm}-1\pm\sqrt{({\rm Pm}-1)^2-4{\rm Pm}{\rm Ha^*}({\rm Ha^*}-2\sqrt{\rm Pm}){\rm Re^*}}}{2\sqrt{\rm Pm}},\nn\\
\lambda_{3,4}&=&\frac{-{\rm Pm}-1\pm\sqrt{({\rm Pm}-1)^2-4{\rm Pm}{\rm Ha^*}({\rm Ha^*}+2\sqrt{\rm Pm}){\rm Re^*}}}{2\sqrt{\rm Pm}}.
\ea
The eigenvalues \rf{mp2} are real near the origin, because ${\rm Ro}=-1$
\ba{mp3}
\lambda_{1,3}&=&-\sqrt{\rm Pm}-{\rm Ha^*}^2\sqrt{\rm Pm}\pm2{\rm Re^*}{\rm Ha^*}{\rm Pm}+o({\rm Pm}), \nn\\
\lambda_{2,4}&=&-\frac{1}{\sqrt{\rm Pm}}+{\rm Ha^*}^2\sqrt{\rm Pm}\mp2{\rm Re^*}{\rm Ha^*}{\rm Pm}+o(\rm Pm).
\ea
Equating the first of the equations \rf{mp2} or \rf{mp3} to zero we reproduce the expression for the threshold \rf{mp0}
at ${\rm Ro}=-1$.

In another particular case, when $\beta^*=0$, ${\rm Re^*}=0$, and ${\rm Ro}=0$ the exact solution to the dispersion equation is two double eigenvalues
$$
\lambda=-\frac{1}{2}\left(\sqrt{\rm Pm}+\frac{1}{\sqrt{\rm Pm}} \right)\pm\sqrt{\frac{1}{4}\left(\sqrt{\rm Pm}-\frac{1}{\sqrt{\rm Pm}} \right)^2-{\rm Ha^*}^2}
$$
that are expressed in terms of the viscous, resistive, and Alfv\'en frequencies in \rf{w6c}.

Consider the eigenvalues corresponding to the values of parameters of Figure \ref{fig9}. Since ${\rm Ro}=-0.85>-1$,
there are two real and two complex branches of eigenvalues $\lambda=\lambda({\rm Pm})$ at $\beta^*=0$ ,
see Figure \ref{fig11a}(a). One of the real branches that comes from minus infinity changes its sign at the threshold \rf{mp0} and excites SMRI with zero eigenfrequency.
Due to the large negative values of the real eigenvalues of the critical branch, there is no way to destabilize the flow at small $\rm Pm$.
However, new opportunities for destabilization occur with the increase of $\beta^*$ that is accompanied by the
qualitative change in the configuration of the eigenvalue branches.

When the azimuthal component of the magnetic field
is switched on, the real eigenvalues of the critical branch get
complex increments, Figure \ref{fig11a}(b,c). With the increase of $\beta^*$,
this critical real branch deforms and                                                
interacts with a stable complex one of an inertial wave until
at $\beta^*\simeq 0.084$ they merge at a point with the origination of the double complex eigenvalue
with the Jordan block known as an exceptional point or EP \cite{B04,MKS05},
see Figure \ref{fig11a}(b).
Notice that another exceptional point corresponds to negative $\beta^*$.
With the further increase of $\beta^*$ this configuration bifurcates into a new one,
where parts of the stable and unstable branches are interchanged, Figure \ref{fig11a}(c). The new critical eigenvalue branch consists of complex eigenvalues that  demonstrate the
typical generalized crossing scenario near an EP \cite{O91,KKM03}, when real parts avoid crossing
while imaginary ones cross and vice versa Figure \ref{fig11a}(c).

In Figure \ref{fig11b} the ''surgery'' of eigenvalue branches is clearly
seen in the complex $({\rm Im}\lambda, {\rm Re}\lambda)$--plane.
Although ''on the surface'' (for ${\rm Re}(\lambda)>0$)
nothing special happens, the deep reason for the
exchange of the fragments between the branches
is ''hidden'' in the ${\rm Re}(\lambda)<0$ region at some
finite value of $\rm Pm$
where an EP is formed, see Figure \ref{fig11b}(c).                   
The critical branch that was responsible for SMRI                   
leaves its stable ''tail'' coming from minus infinity (Figure \ref{fig11b}(a,b))     
and instead ''catches'' a fragment of a
stable branch of complex eigenvalues with small real parts that comes from
zero (Figure \ref{fig11b}(d-f)). This re-arranged                           
branch of complex eigenvalues is much more prone to instabilities at
low $\rm Pm$ than the original critical one as the further increase
of $\beta^*$ confirms. Indeed, the negative real parts become
smaller and around $\beta^*=0.7$ there appears a new interval of
HMRI at those values of $\rm Pm$, at which SMRI did not exist, see
Figure \ref{fig11c}(a). This interval exactly corresponds to the
island of HMRI shown in Figure \ref{fig9}(b,c). We note here that
the numerical calculation of the roots of the dispersion relation
confirms the boundaries of the regions of HMRI given by the Bilharz
criterion: $m_4=0$.

The hidden exceptional point governs transfer of instability
between the  branch of (helically modified) SMRI and a               
complex branch of the inertial wave that after interaction
becomes prone to destabilization.
This qualitative effect explains why switching the azimuthal
component of the magnetic field on
we get HMRI as a travelling wave whereas SMRI was a stationary        
instability.
Moreover, as Figure \ref{fig11c}(c) shows, the new critical branch
is characterized by a broad band of unstable frequencies while
the tail of the branch
responsible for SMRI corresponds to a more sharply selected
unstable frequency which is close to zero at $\beta^*\ne 0$.

The above observations are in agreement with the observation of Liu
et al. (2006) that, in contrast to SMRI, which is a destabilized
slow Magneto-Coriolis wave, HMRI is a weakly destabilized inertial
oscillation. Further results on interpretation of the HMRI as an
unstable MHD-wave as well as on its relation to the
dissipation-induced instabilities will be published elsewhere
\cite{FKS10}.

\subsubsection{Second scenario of HMRI excitation}

The remarkable complexity of the phenomenon of HMRI manifests
itself in different scenarios of destabilization.
It turns out, that the transition from SMRI to HMRI through
the exceptional point is not the only way to instabilities
at low magnetic Prandtl numbers. At higher values of        
${\rm Re}^*$ and ${\rm Ha}^*$, in the presence of the
azimuthal magnetic field the inertial wave can become          
unstable without mixing with the critical SMRI branch.

In contrast to the scenario of the first type when one mixed
complex branch becomes unstable at different
intervals of ${\rm Pm}$ and causes both the essential HMRI
and the helically modified SMRI,
in the new situation the inertial wave causes the excitation
of the essential HMRI
and the critical real branch remains responsible for the
helically modified SMRI, as is clearly seen in Figure \ref{fig12}.

Most surprisingly, the inertial wave branch can become             
unstable twice with the increase of $\beta^*$.
The first time this happens in the vicinity of                     
$\beta_1^*=\frac{1}{2}\sqrt{2}\simeq 0.707$, see Figure \ref{fig12}(b),
then---in the neighborhood
of $\beta_2^*=\frac{3}{2}\sqrt{2}\simeq 2.121$, as is visible in Figure \ref{fig12}(c).
In the $({\rm Pm},\beta^*)$-plane this yields two islands of the essential HMRI that coexist with the continent of
the helically modified SMRI, Figure \ref{fig13}(a). The
real parts of the unstable branches shown in black and grey in
Figure \ref{fig16} correspond to the first and second
essential HMRI islands and to the continent of the helically modified
SMRI, respectively.

The difference between the two HMRI scenarios is visible also in the
other parameter planes. For example, in $({\rm Ha}^*,\rm Ro)$-plane
the domain of HMRI developed by the first scenario and corresponding
to the HMRI-island of Figure \ref{fig13}(c) has an SMRI-like form
with two peaks, see Figure \ref{fig13}(d). The second HMRI
excitation scenario leads to the inclusions of the essential HMRI in
the helically modified SMRI domain, Figure \ref{fig13}(b). Although
the islands in Figure \ref{fig13}(a) and Figure \ref{fig13}(c) look
similar, the former are slices of the two three-dimensional domains
that intersect each other along an edge, while the latter are slices
of the same smooth three-dimensional region of instability, cf.
Figure \ref{fig14}(a).

\section{Conclusions}

The helical magnetorotational instability is a more complicated
phenomenon than the standard one.
We found evidences that HMRI can be identified with the destabilization
of an inertial                                                            
wave in contrast to SMRI that is a destabilized slow Magneto-Coriolis
wave. We established two scenarios of transition
from SMRI to HMRI: the first one is accompanied by the origination
of a spectral exceptional point and a transfer of
instability between modes while in the second scenario two independent
eigenvalue branches become unstable.
We distinguish between the essential HMRI that is characterized
by small magnetic Prandtl numbers at which SMRI is not possible, smaller
growth rates than SMRI, and by non-zero
frequencies and the helically modified SMRI which is caused by a small
perturbation of the unstable real eigenvalue
branch and is thus characterized by high growth rates, small frequency
and relatively high magnetic Prandtl numbers
within the usual range of SMRI. With the use of the Bilharz stability
criterion we established explicit expressions for
the stability boundary and proved rigorously the bounds on the
critical Rossby number
for HMRI in the inductionless limit $(\rm Pm=0)$.
Nevertheless, we revealed that for $\rm Pm \ne 0$ these bounds
can be easily exceeded---an indicator
in favor of the HMRI for small negative Rossby numbers. Finally,
we found that for small negative Rossby numbers
the essential HMRI forms separated islands that can coexist
simultaneously in the $({\rm Pm},\beta^*)$-plane.

\acknowledgments
The work of O.N.K. has been supported by the research grant DFG HA 1060/43-1.
The work of F.S. has been supported by the German Leibniz Gemeinschaft
within its SAW programme. F.S. likes also to thank G. Gerbeth, R. Hollerbach,
J. Priede and G. R\"udiger for many stimulating discussions.





\appendix

\section{Linearization with respect to non-axisymmetric perturbations}

We linearize equations \rf{m1}-\rf{m3} in the vicinity of the stationary solution \rf{m4}-\rf{m4c} assuming general perturbations ${\bf u}={\bf u}_0+{\bf u}'$, $p=p_0+p'$, and ${\bf B}={\bf B}_0+{\bf B}'$ and leaving only the terms of first order with respect to the primed quantities. With the notation \cite{G02,LGHJ06}
\be{l0}
\partial_t=\frac{\partial}{\partial t},\quad \partial_{\phi}=\frac{\partial}{\partial \phi},\quad \partial_R=\frac{\partial}{\partial R},\quad \partial_z=\frac{\partial}{\partial z}, \quad \partial_R^{\dagger} = \partial_R +\frac{1}{R},\quad D= \partial_R^{\dagger}\partial_R+\frac{1}{R^2}\partial^2_{\phi}+\partial_z^2
\ee
we write the linearized equations in cylindrical coordinates, cf. \cite{G02,PP05,LGHJ06}
\ba{l1}
(\partial_t-\nu D) u_R'-2\Omega u_{\phi}'+\Omega\partial_{\phi}u_R'
&=&
-\frac{1}{\rho}\left[\partial_R p'+\frac{1}{\mu_0}\left(B_z^0{\partial_R}B_z' + B_{\phi}^0{\partial_R}B_{\phi}'-\frac{B_{\phi}^0}{R}B_{\phi}'\right)\right]\nn\\
&+&
\frac{1}{\mu_0 \rho} \left( B_z^0 {\partial_z}B_R'+\frac{B_{\phi}^0}{R}\partial_{\phi} B_R' -2\frac{B_{\phi}^0}{R} B_{\phi}'\right)\nn\\&-&\nu\left(\frac{u_R'}{R^2}+\frac{2}{R^2}{\partial_{\phi} u_{\phi}'}\right),\nn\\
({\partial_t}-\nu D)u_{\phi}'+\frac{\kappa^2}{2\Omega}u_R'+\Omega{\partial_{\phi} u_{\phi}'}&=&-\frac{1}{\rho}\frac{1}{R}\left[{\partial_{\phi} p'}+\frac{1}{\mu_0}\left(B_z^0{\partial_{\phi} B_z'} + B_{\phi}^0{\partial_{\phi} B_{\phi}'}\right)\right]\nn\\
&+&\frac{1}{\mu_0\rho} \left(B_z^0{\partial_z B_{\phi}'}+\frac{B_{\phi}^0}{R}{\partial_{\phi} B_{\phi}'}\right)+\nu\left(\frac{2}{R^2}{\partial_{\phi} u_R'}-\frac{u_{\phi}'}{R^2}\right),\nn\\
({\partial_t}-\nu D)u_z'+\Omega{\partial_{\phi} u_z'}&=&
-\frac{1}{\rho}\left[{\partial_z p'}+\frac{1}{\mu_0}\left(B_z^0{\partial_z B_z'} + B_{\phi}^0{\partial_z B_{\phi}'}\right)\right]\nn\\&+&\frac{1}{\mu_0\rho} \left(B_z^0{\partial_z B_z'}+\frac{B_{\phi}^0}{R}{\partial_{\phi} B_z'}\right),\nn\\
({\partial_t}-\eta D)B_R'&=&\frac{B_{\phi}^0}{R}{\partial_{\phi} u_R'}-\Omega{\partial_{\phi} B_{\phi}'}+B_z^0 {\partial_z u_R'}
-\eta\left(\frac{B_R'}{R^2}+\frac{2}{R^2}{\partial_{\phi} B_{\phi}'} \right),\nn\\
({\partial_t}-\eta D )B_{\phi}'&=&B_z^0{\partial_z u_{\phi}'}-B_{\phi}^0{\partial_z u_z'}-
B_{\phi}^0{\partial_R u_R'}-u_R'{\partial_R B_{\phi}^0}\nn\\
&+&u_{\phi}^0 {\partial_R B_R'}+u_{\phi}^0 {\partial_z B_z'}+B_R'{\partial_R u_{\phi}^0}\nn\\
&+&\eta\left(
\frac{2}{R^2}{\partial_{\phi} B_R'}-\frac{B_{\phi}'}{R^2}\right),\nn\\
({\partial_t}-\eta D)B_z'&=&
-B_z^0 {\partial_R u_R'}-\frac{B_z^0}{R}u_R'-\frac{B_z^0}{R}{\partial_{\phi} u_{\phi}'}-
\Omega{\partial_{\phi} B_z'}+\frac{B_{\phi}^0}{R}{\partial_{\phi} u_z'},\nn\\
0&=&\partial_R^{\dagger} u_R'+\frac{1}{R}{\partial_{\phi} u_{\phi}'}+{\partial_z u_z'},\nn\\
0&=&\partial_R^{\dagger} B_R'+\frac{1}{R}{\partial_{\phi} B_{\phi}'}+{\partial_z B_z'}.
\ea

\newpage

\newpage

\newpage

\begin{figure}
\includegraphics[width=160mm]{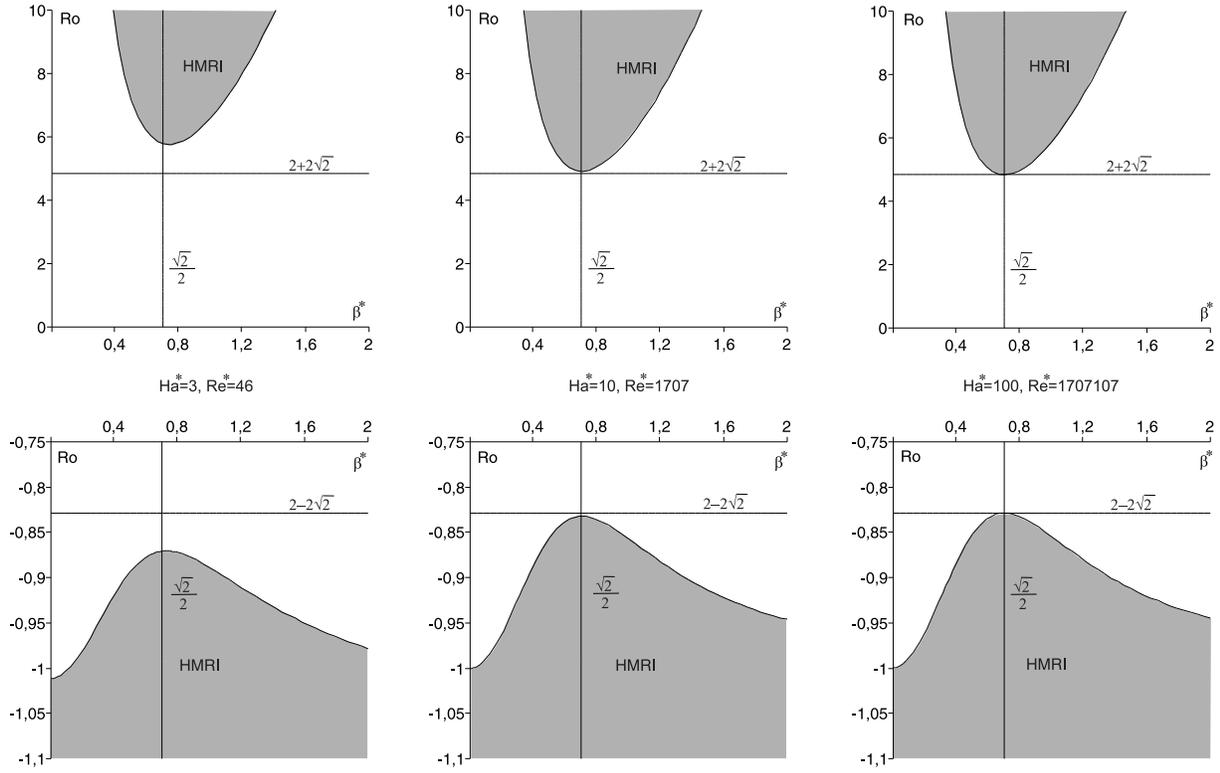}
\caption{\label{fig3n} Inductionless approximation $(\rm Pm=0)$: Two
domains \rf{a2} of HMRI (grey) are within the bounds \rf{i11}. It is
seen how the scaling law \rf{i5} works. HMRI is possible even for
$\rm Pm=0$ -- the paradox of inductionless magnetorotational
instability \cite{PGG07}.  }
\end{figure}

\newpage

\begin{figure}
\includegraphics[width=160mm]{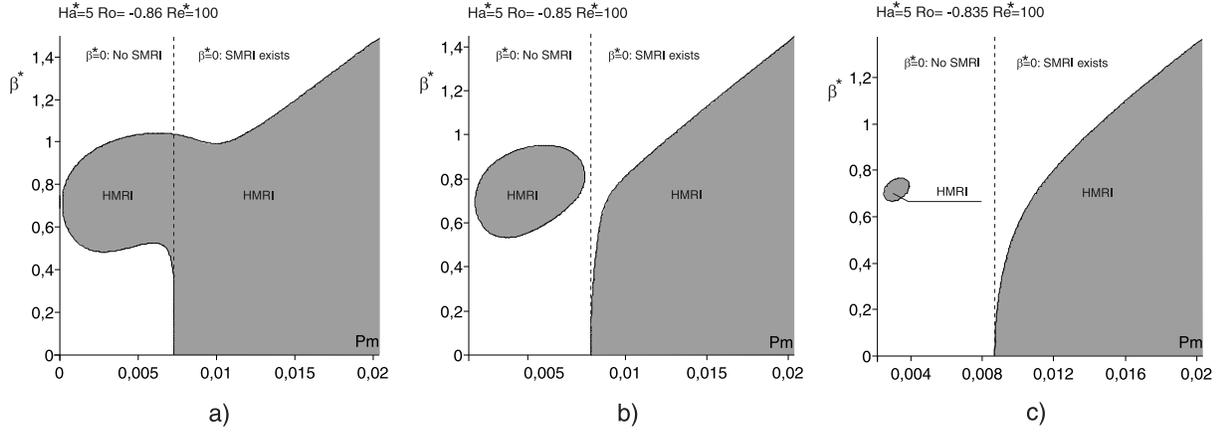}
\caption{\label{fig9} The relation of SMRI (right to the dashed line
for $\beta^*=0$) and HMRI (grey area). For $\beta^*\ne 0$ a
(semi)-island of HMRI exists in the neighborhood of
$\beta^*=\sqrt{2}/2$ for the values of $\rm Pm$ at which SMRI was
not possible in the absence of the azimuthal magnetic field
$(\beta^*=0)$. The size of the HMRI island decreases with the
increase of $\rm Ro$. }
\end{figure}

\newpage

\begin{figure}
\includegraphics[width=160mm]{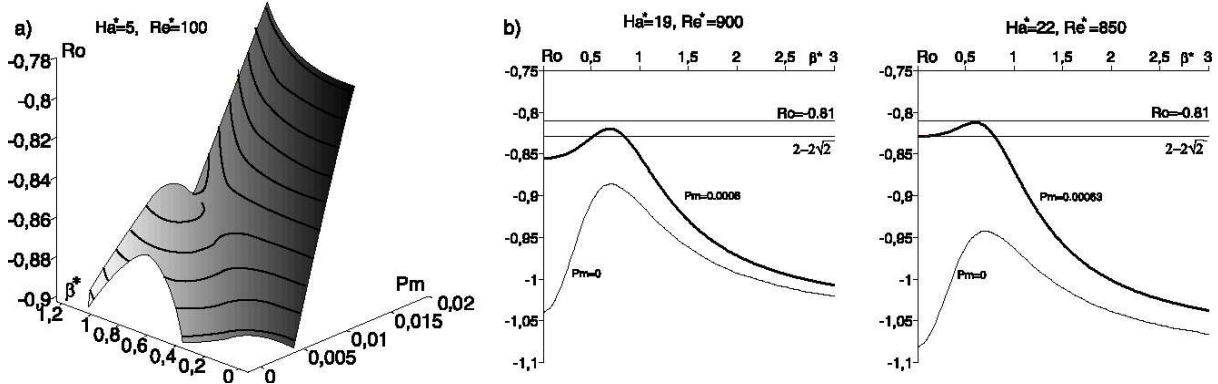}
\caption{\label{fig14} (a) The three-dimensional HMRI            
domain shows how the slices of it in the plane ${\rm Pm} = const$ converge smoothly to the inductionless HMRI domain for $\beta^*=0$. (b,c) HMRI domain ($m_4>0$) can
exceed the inductionless bound $\rm Ro = 2-2\sqrt{2}$ when $\rm Pm \ne 0$.
Thin lines show the boundary of the inductionless                     
HMRI domain, bold ones show that with $\rm Pm \ne 0$. }               
\end{figure}

\newpage

\begin{figure}
\includegraphics[width=160mm]{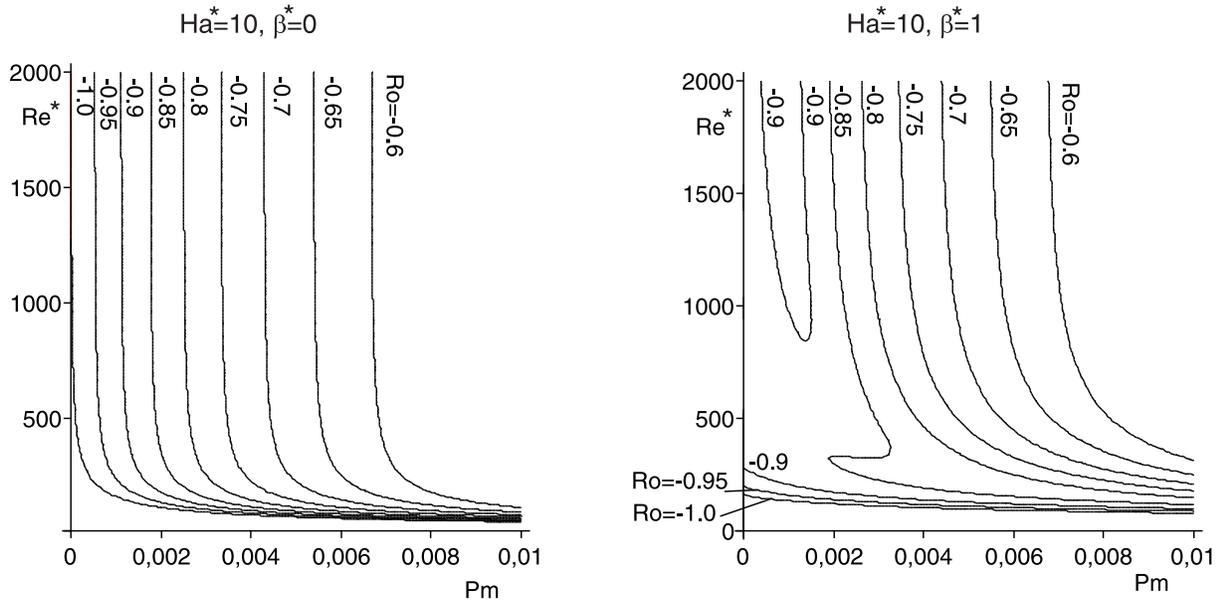}
\caption{\label{fig15} SMRI boundaries and those of HMRI in the
plane (${\rm Pm}, {\rm Re}^*$) for various values of $\rm Ro$. A
considerable deformation of the boundaries when $\beta^*$ is
switched on is seen in the region of small $\rm Pm$.  The pictures
are in agreement with the calculations of \cite{RS08}.}
\end{figure}

\newpage

\begin{figure}
\includegraphics[width=160mm]{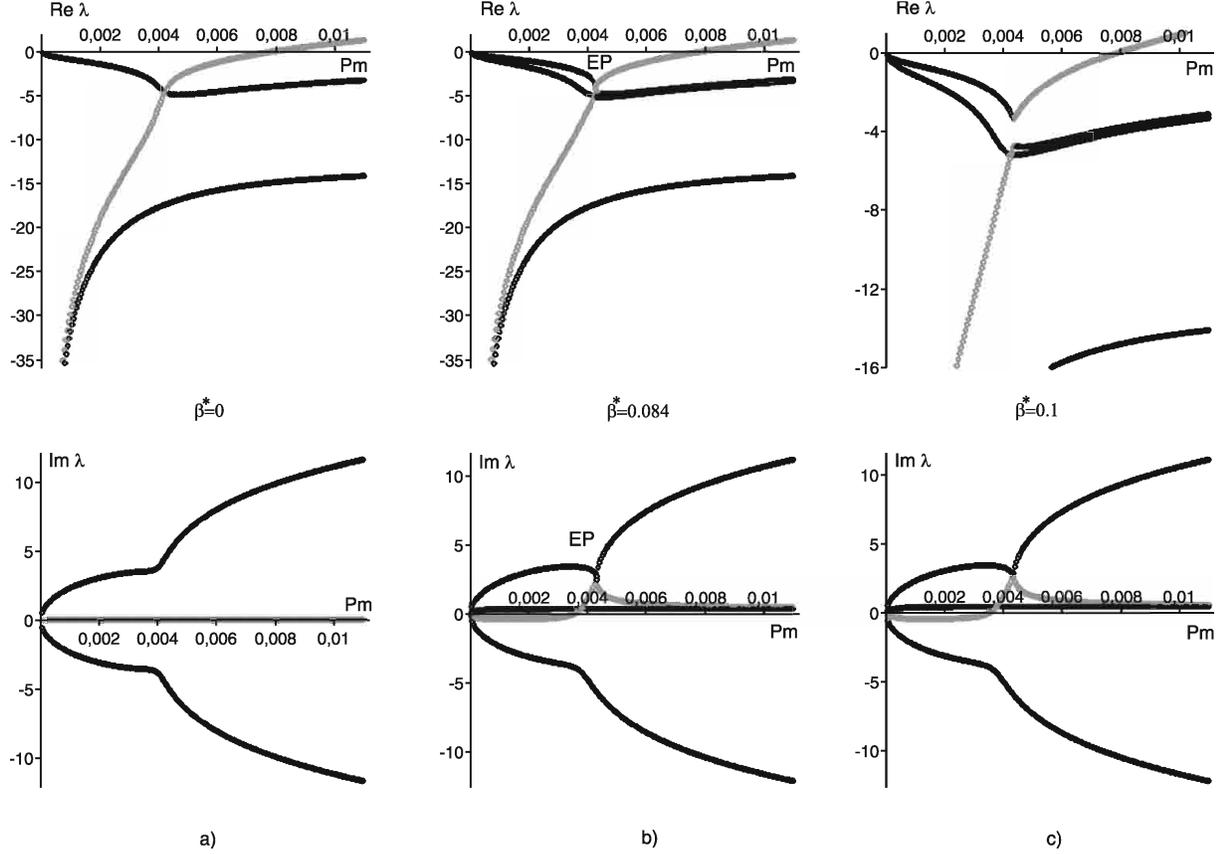}
\caption{\label{fig11a} ${\rm Ha}^*=5$, ${\rm Re}^*=100$, ${\rm
Ro}=-0.85$: Real and imaginary parts of eigenvalues $\lambda$ as
functions of $\rm Pm$. (a) Two stable complex branches (inertial
modes) and two real ones, one of the latter (grey) becomes positive
and causes SMRI; (b) merging of the inertial mode with the deformed
unstable SMRI-branch (grey) with the origination of an exceptional
point (EP); (c) bifurcation yields new mixed eigenvalue branches
that create new opportunities for instability at low $\rm Pm$.}
\end{figure}

\newpage

\begin{figure}
\includegraphics[width=160mm]{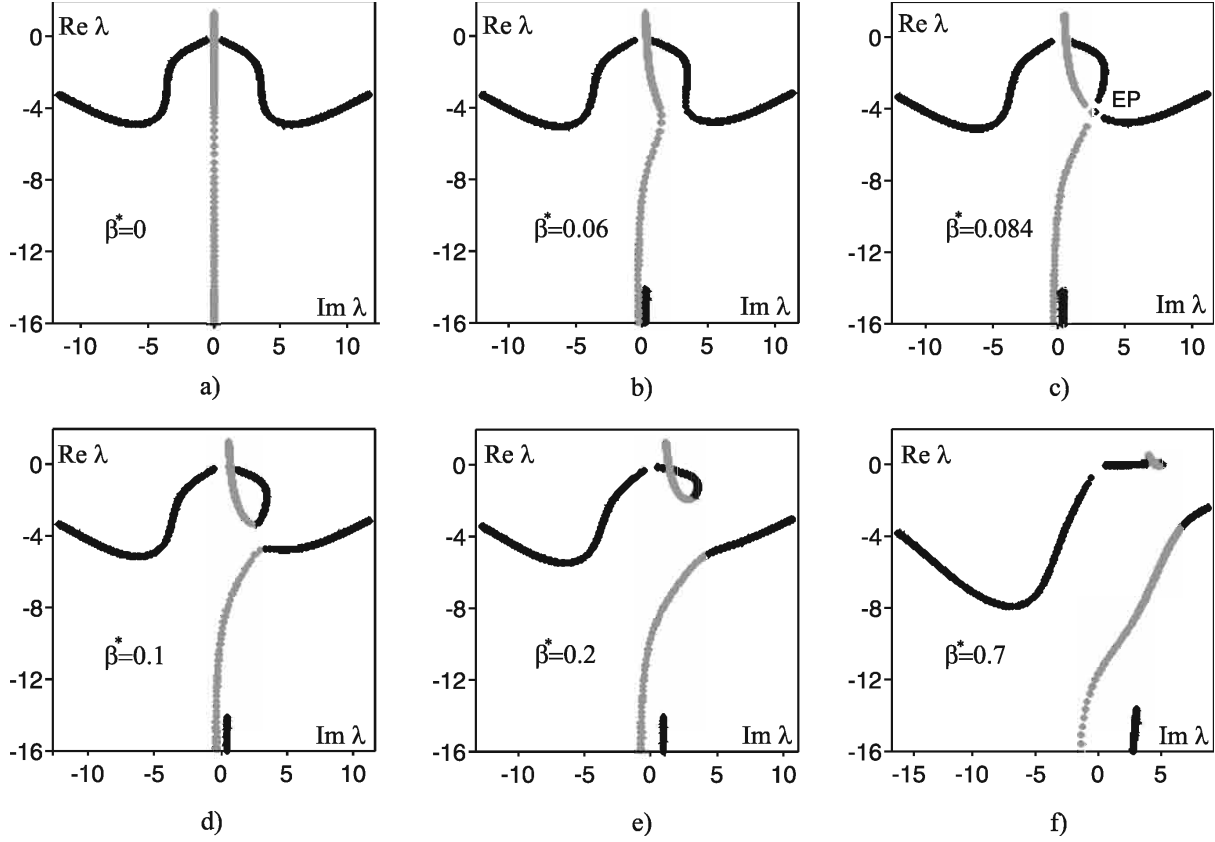}
\caption{\label{fig11b} ${\rm Ha}^*=5$, ${\rm Re}^*=100$, ${\rm
Ro}=-0.85$: Eigenvalue trajectories in the complex plane when $\rm
Pm$ changes from zero to ${\rm Pm}=0.011$ show the effect of
transfer of instability between branches through a spectral
exceptional point (EP). (a) SMRI is caused by the transition of a
pure real eigenvalue from negative values to positive. (b)
deformation of the critical branch (grey) for $\beta^*\ne 0$; (c)
merging of the critical and stable branches with the origination of
the double complex eigenvalue with the Jordan block - exceptional
point (EP); (d-f) splitting of the EP and bifurcation of the
eigenvalue trajectories: after a surgery the stable complex branch
acquires an unstable tail of the critical branch (grey) and becomes
unstable near the origin, which corresponds to an unstable traveling
wave of HMRI.}
\end{figure}

\newpage

\begin{figure}
\includegraphics[width=160mm]{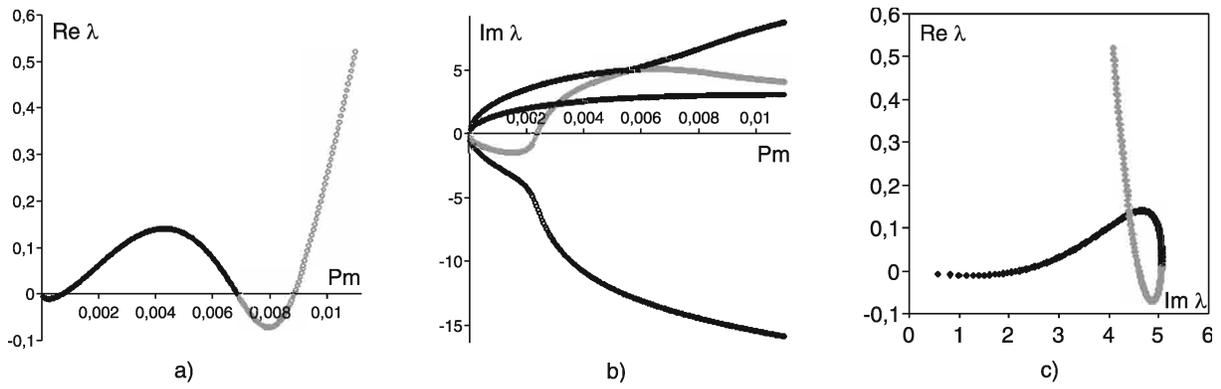}
\caption{\label{fig11c} ${\rm Ha}^*=5$, ${\rm Re}^*=100$, ${\rm
Ro}=-0.85$, $\beta^*=0.7$: (a) Real part of the critical eigenvalue
branch is positive while $\rm Pm$ is within the island or the
continent of instability; (b) imaginary parts of stable and unstable
branches; (c) trajectory of the critical eigenvalue that creates
instability with the change of $\rm Pm$ from zero to ${\rm
Pm}=0.011$.}
\end{figure}

\newpage

\begin{figure}
\includegraphics[width=160mm]{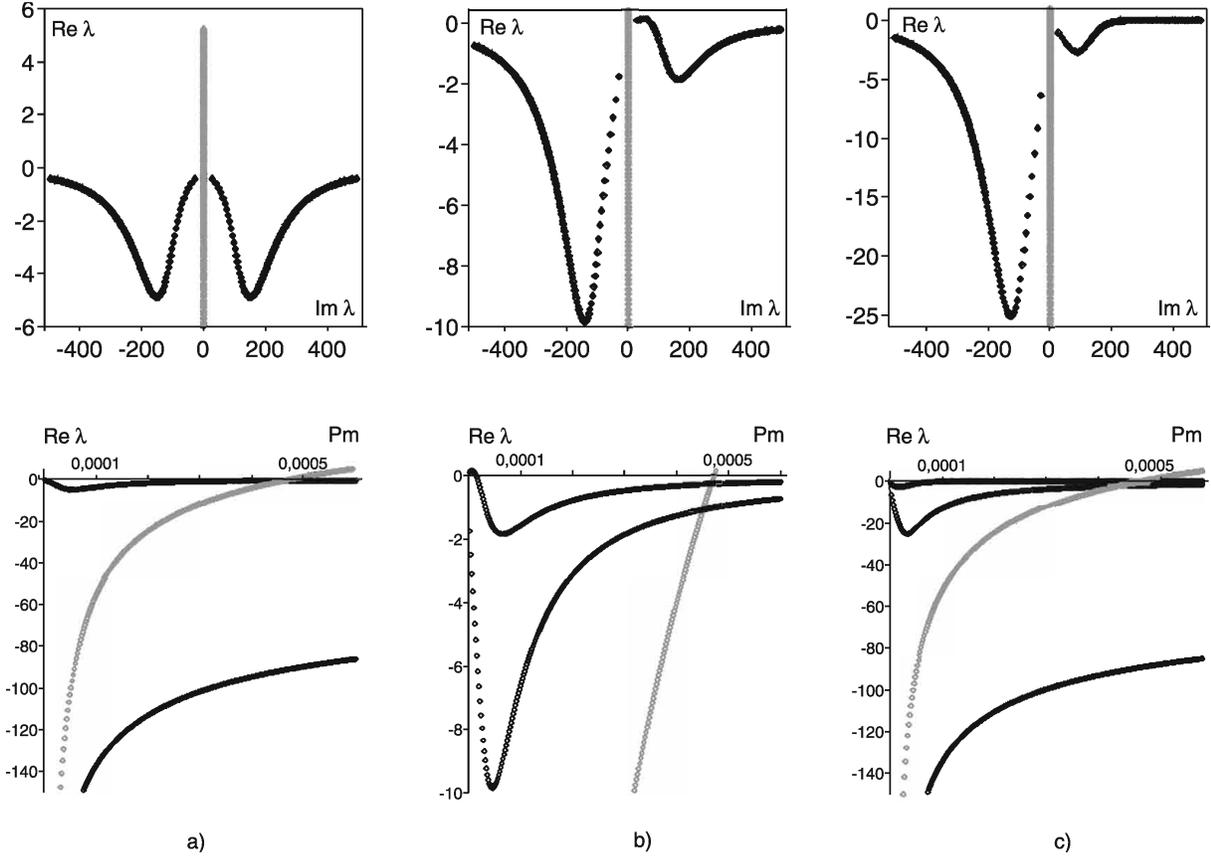}
\caption{\label{fig12} ${\rm Ha}^*=17$, ${\rm Re}^*=28850$, ${\rm
Ro}=-0.88$: (a) Stable inertial waves (black) and critical SMRI
branch of real eigenvalues (grey) for $\beta^*=0$; (b) for
$\beta^*=0.7$ one of the inertial wave branches becomes unstable
independently of the SMRI branch; (c) for $\beta^*=2$ the same
inertial wave branch becomes unstable a second time at higher
frequencies independently of the SMRI branch.}
\end{figure}

\newpage

\begin{figure}
\includegraphics[width=160mm]{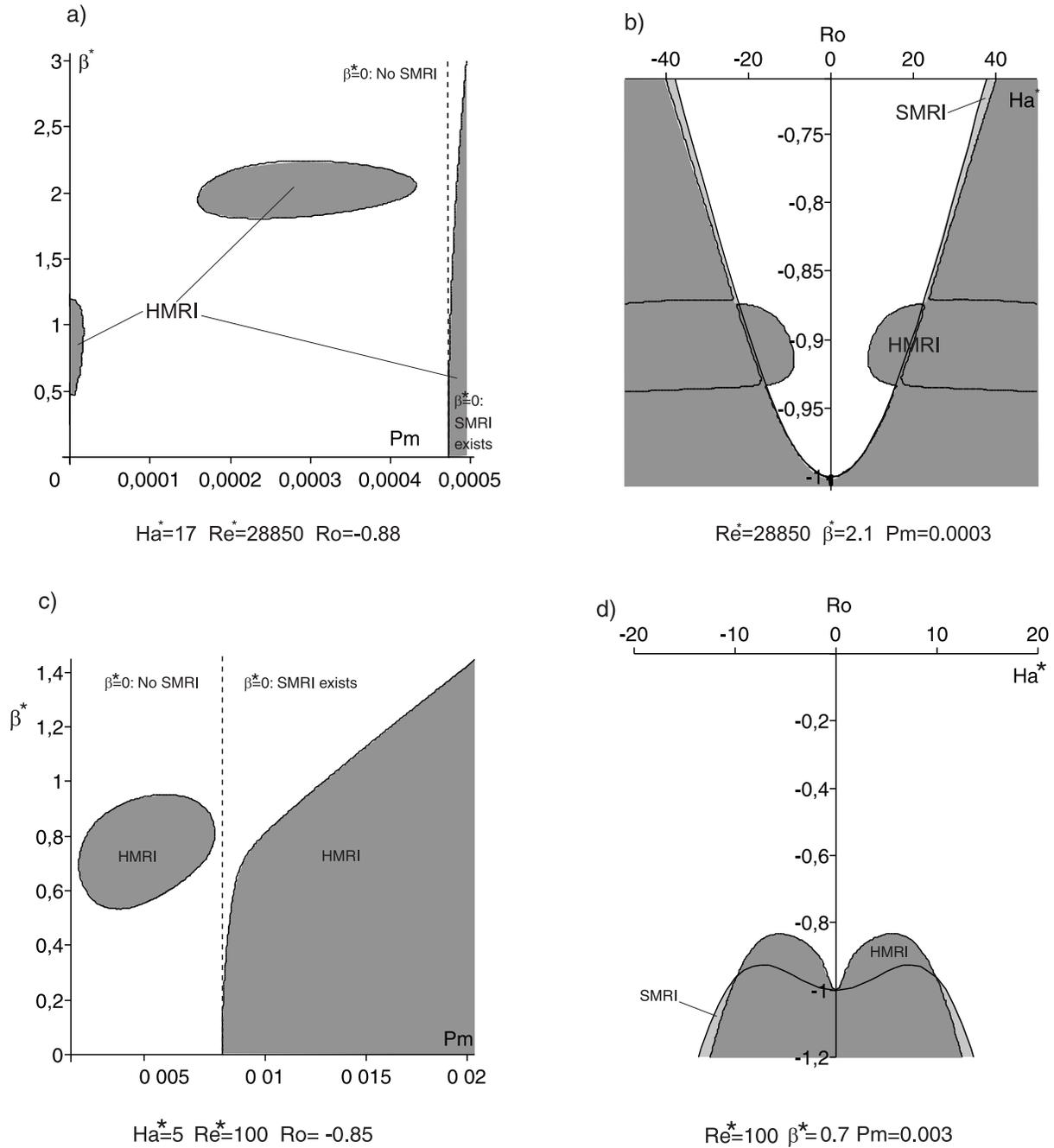}
\caption{\label{fig13} (a) The first and second islands of the
essential HMRI developed according to the second scenario in the
neighborhood of $\beta^*=\sqrt{2}/2$ and $\beta^*=3\sqrt{2}/2$, (b)
an essential HMRI inclusion in the helically modified SMRI domain,
(c) an island of the essential HMRI developed according to the first
scenario, (d) the corresponding view in the $({\rm Ha}^*,{\rm
Ro})$-plane. }
\end{figure}

\newpage

\begin{figure}
\includegraphics[width=160mm]{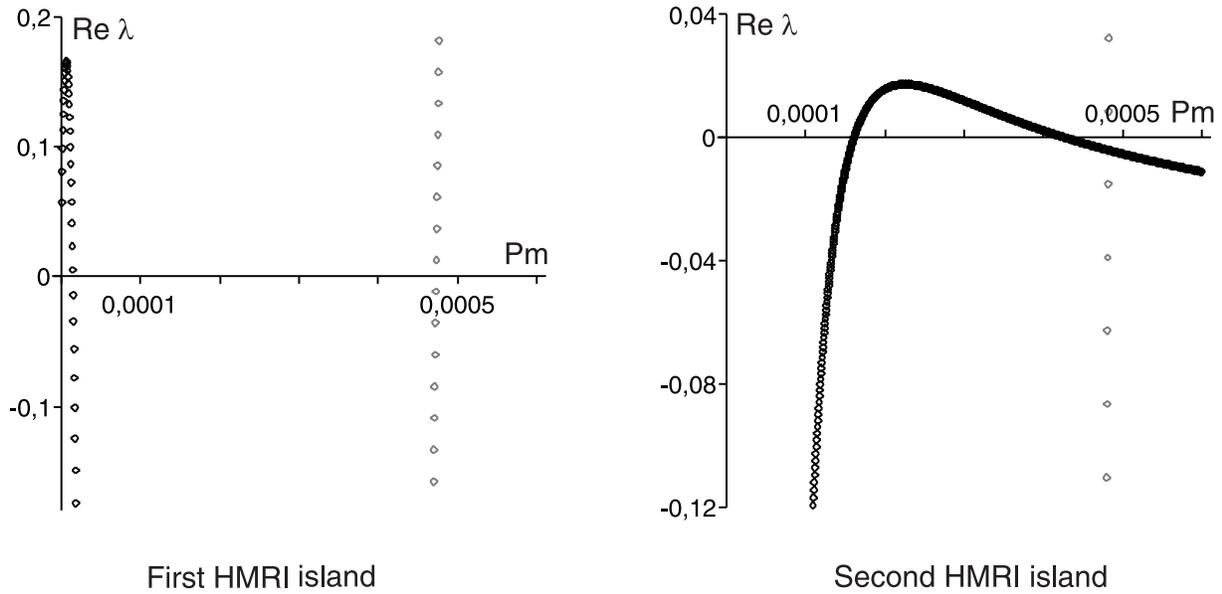}
\caption{\label{fig16} ${\rm Ha}^*=17$, ${\rm Re}^*=28850$, ${\rm
Ro}=-0.88$: (left, $\beta^*=0.7$) Real part of the first critical
eigenvalue branch (grey) is positive when $\rm Pm$ is within the
continent while for the second critical branch (black) it is within
the first HMRI island; (right, $\beta^*=2$) real part of the first
critical eigenvalue branch (grey) is positive when $\rm Pm$ is
within the continent while for the second critical branch (black) it
is within the second HMRI island.}
\end{figure}

\end{document}